\documentclass[a4paper,12pt]{article}
\pdfoutput=1
\usepackage{graphicx}
\usepackage{color}
\usepackage{latexsym}
\usepackage{amssymb,euscript, mathrsfs}
\usepackage{braket}
\usepackage{mathtools}
\usepackage{amsmath}
\usepackage{bbm} %for \mathbbm{1}
\usepackage{verbatim}

\usepackage{jheppub}                   % jhep style
\makeatletter
\gdef\@fpheader{\ }                    % hack the jhep header
\makeatother

\newcommand{\CA}{\mathcal{A}}
\newcommand{\CB}{\mathcal{B}}

\newcommand{\CO}{\mathcal{O}}

\newcommand{\CV}{\mathcal{V}}
\newcommand{\CS}{\mathcal{S}}

\newcommand{\CW}{\mathcal{W}}

\def\be{\begin{equation}}
\def\ee{\end{equation}}

\newcommand{\nn}{\nonumber}
\newcommand{\diff}{\mathrm{d}}
\newcommand{\SO}{{\rm SO}}
\newcommand{\SU}{{\rm SU}}

\newcommand{\U}{{\rm U}}

\newcommand{\air}{\alpha}
\newcommand{\hir}{\eta}
\newcommand{\HL}{\widetilde{H}}

\title{Squashing the Boundary of Supersymmetric AdS$_{\mathbf 5}$ Black Holes}

\author[a]{Davide Cassani,}
\emailAdd{davide.cassani@pd.infn.it}
\author[\,a,b]{Lorenzo Papini}
\emailAdd{lorenzo.papini@pd.infn.it}

\affiliation[a]{INFN, Sezione di Padova, Via Marzolo 8, 35131 Padova, Italy}

\affiliation[b]{Dipartimento di Fisica e Astronomia ``Galileo Galilei'', Via Marzolo 8, 35131 Padova, Italy}

\abstract{We construct new supersymmetric black holes in five-dimensional supergravity with an arbitrary number of vector multiplets and Fayet-Iliopoulos gauging. These are asymptotically locally AdS$_5$ and the conformal boundary comprises a squashed three-sphere with $\SU(2)\times\U(1)$ symmetry.
The solution depends on two parameters, of which one determines the angular momentum and the Page electric charges, while the other controls the squashing at the boundary. The latter is arbitrary, however in the flow towards the horizon it is attracted to a value that only depends on the other parameter of the solution. The entropy is reproduced by a simple formula involving the angular momentum and the Page charges, rather than the holographic charges. Choosing the appropriate five-dimensional framework, the solution can be uplifted to type IIB supergravity on $S^5$ and should thus be dual to $\mathcal{N}=4$ super Yang-Mills on a rotating, squashed Einstein universe.}

\begin{document}
\maketitle

%%%%%%%%%%%%%%%%%%%%%%%%%%%%%%%%%%%%%%%%%%%%%%%%%%%%%%%%%%%%%%%%%%%%%%%%%%%%%%%%%%%%%%
\section{Introduction}

A main challenge in quantum gravity is to explain the black hole entropy via a microstate counting. For extremal black holes one can attack this problem using the AdS$_2$/CFT$_1$ correspondence \cite{Sen:2008vm}, however in most cases the relevant conformal quantum mechanics is not known and it is therefore hard to compute the ground state degeneracy that should account for the black hole entropy precisely. When the black hole is asymptotic to AdS$_{d+1}$ (with $d>1$), one can hope to exploit the additional leverage of the AdS$_{d+1}$/CFT$_d$ correspondence, which in many instances is well under control. Recently this strategy has led to exciting results for supersymmetric black holes in AdS$_4$. In \cite{Benini:2015eyy,Benini:2016rke}, the classical entropy of a class of static, dyonically charged supersymmetric AdS$_4$ black holes with an uplift to M-theory on $S^7$ has been reproduced by evaluating the large $N$ limit of a suitably defined partition function of the ABJM superconformal field theory on $S^1\times S^2$. This result has then been extended to further examples of AdS$_4$ black holes in M-theory and massive type IIA string theory e.g.\ in \cite{Hosseini:2016tor,Benini:2016hjo,Cabo-Bizet:2017jsl,Azzurli:2017kxo,Hosseini:2017fjo,Benini:2017oxt}, while subleading corrections in the large $N$ expansion have been investigated for example in \cite{Liu:2017vbl}.

Supersymmetric asymptotically AdS$_5$ black holes with an uplift to string theory have been known for some time \cite{Gutowski:2004ez,Gutowski:2004yv,Chong:2005hr,Chong:2005da,Kunduri:2006ek}, however the attempts to match their entropy via a four-dimensional field theory computation have not been equally satisfying so far \cite{Kinney:2005ej}. Very recently an interesting observation has been made \cite{Hosseini:2017mds}, that the entropy of the known supersymmetric AdS$_5$ black holes is reproduced by extremizing a quantity which appears to be closely related to the supersymmetric Casimir energy of four-dimensional superconformal field theories (SCFT's) on $S^1 \times S^3$ \cite{Assel:2014paa,Assel:2015nca,Bobev:2015kza}.
New hairy black holes with the same AdS$_5$ asymptotics have also been proposed and put in the context of the entropy puzzle \cite{Markeviciute:2018yal}.

Further information on the field theory states that contribute to the entropy might come from studying whether the black hole solutions continue to exist when one tries to deform the geometry of the conformal boundary, and if so how this affects their thermodynamics. This question has recently been investigated in \cite{Blazquez-Salcedo:2017kig,Blazquez-Salcedo:2017ghg}.\footnote{See also \cite{Murata:2009jt} for a non-supersymmetric study.} Working in minimal five-dimensional gauged supergravity and using a cohomogeneity-one ansatz with local $\SU(2) \times \U(1)\times \U(1)$ symmetry, the authors constructed both supersymmetric and non-supersymmetric black holes where the three-sphere sitting at the conformal boundary of global AdS$_5$ is squashed. Since the boundary is non conformally-flat, the solutions are only Asymptotically {\it locally} AdS$_5$ (AlAdS$_5$), rather than asymptotically AdS$_5$.  
While the squashing at the boundary is arbitrary, in the supersymmetric case the event horizon geometry turns out to be completely frozen and therefore the entropy takes a fixed value. This behaviour is qualitatively different from the one of asymptotically AdS$_5$ black hole solutions to minimal gauged supergravity with the same symmetry \cite{Gutowski:2004ez}, where the entropy depends on one parameter controlling the horizon geometry.

Motivated by the above developments, in this paper we construct more general supersymmetric black holes having a local $\SU(2) \times \U(1)\times \U(1)$ symmetry 
 and displaying a squashed three-sphere at the boundary.
We address this problem in the context of five-dimensional Fayet-Iliopoulos gauged supergravity. This is five-dimensional supergravity coupled to an arbitrary number $n_V$ of vector multiplets and with a $\U(1)$ gauging of the R-symmetry. It is expected to describe holographically a subsector of dual $\mathcal{N}=1$ SCFT's, made of the $\mathcal{N}=1$ energy-momentum tensor multiplet and $\U(1)^{n_V}$ flavour current multiplets. This can be made rigorous by focusing on a specific model with $n_V=2$, which is a consistent truncation of type IIB supergravity on $S^5$ and is thus dual to $\mathcal{N}=4$ super-Yang-Mills (seen as an $\mathcal{N}=1$ theory). However we can work in more generality and keep $n_V$ arbitrary in our discussion.

In the solutions we will look for, one of the Abelian Killing vectors is timelike while the remaining $\SU(2)\times\U(1)$ symmetry acts on a three-sphere.  
The a priori non-vanishing, conserved charges carried by the solutions thus are the energy, one angular momentum and $n_V+1$ electric charges.
Previously known supersymmetric solutions with the same symmetry include the black holes of \cite{Gutowski:2004ez,Gutowski:2004yv}, the black hole with a squashed boundary of \cite{Blazquez-Salcedo:2017kig,Blazquez-Salcedo:2017ghg} and a solitonic deformation of AdS found in~\cite{Cassani:2014zwa}. Apart for the solution of \cite{Gutowski:2004yv}, these were obtained by restricting to minimal gauged supergravity and thus have just one electric charge.

The relevant conditions for a supersymmetric solution to Fayet-Iliopoulos gauged supergravity were given in \cite{Gutowski:2004yv}. By partially solving these conditions and imposing an ansatz on the scalar fields, we are able to reduce the problem to two coupled ordinary differential equations. 
However these are very complicated and we could not find new analytic solutions. We rather construct the near-horizon and near-boundary solutions perturbatively and then interpolate numerically.
In this way we obtain a two-parameter family of supersymmetric black holes displaying both running gauge fields and scalar fields and generalizing the one-parameter solution of \cite{Blazquez-Salcedo:2017kig,Blazquez-Salcedo:2017ghg}. We show that for a certain range of the parameters the solution is regular on and outside the horizon.

We find that of the two parameters, one controls the event horizon geometry as well as the angular momentum and the Page electric charges of the solution, while the other is responsible for the squashing at the boundary and does not affect the horizon. This means that whatever is the squashing at the boundary, the radial flow towards the horizon acts as an attractor that brings the transverse geometry into a form which only depends on the other parameter. Still, the horizon is not frozen and the entropy is a non-trivial function of this other parameter. 

We are eventually interested in the holographic interpretation of the solution. By examining the asymptotic modes of the supergravity fields near the conformal boundary we can determine the dual $\mathcal{N}=1$ SCFT background fields. We find that in addition to a squashed three-sphere, the field theory background features non-vanishing field strengths for the non-dynamical gauge fields coupling to the R-current and $n_V$ flavour currents, as well as non-vanishing D-terms sourcing the scalar superpartners of the flavour currents (as required by supersymmetry). We then set up holographic renormalization for Fayet-Iliopoulos gauged supergravity, providing the needed counterterms. This allows to compute the holographic one-point function for the SCFT energy-momentum tensor, R-current, flavour currents and the scalar superpartners of the latter.
These in turn provide the holographic energy, the angular momentum and the R- and flavour charges. While these conserved quantities are naturally interpreted as expectation values of the corresponding SCFT operators in the state dual to the black hole, they also make sense in the gravitational solution, independently of holography.
In addition we compute the renormalized on-shell action and verify that it satisfies the quantum statistical relation.
Finally, we find that the black hole entropy can be expressed as a simple function of the angular momentum and the Page electric charges, but apparently not the holographic electric charges.

The rest of the paper is organized as follows. In Section~\ref{sec:setup} we summarize the essential features of Fayet-Iliopoulos gauged supergravity and elaborate on the supersymmetry conditions. A summary of the resulting equations is given in Section~\ref{SummarySusyEq}. In Section~\ref{sec:Solution} we present our new solution. In Section~\ref{sec:properties_sol} we discuss holographic renormalization in Fayet-Iliopoulos supergravity and apply it to the evaluation of the holographic charges as well as the on-shell action. We also discuss the entropy of the solution. 
 We conclude in Section~\ref{sec:conclusions}. Appendix~\ref{app:conditions_qI} collects some technical details of our computations while Appendix~\ref{app:FeffermanGraham} displays the asymptotic solution in Fefferman-Graham form.

%%%%%%%%%%%%%%%%%%%%%%%%%%%%%%%%%%%%%%%%%%%%%%%%%%%%%%%%%%%%%%%%%%%%%%%%%%%%%%%%%%%%%%
\section{Setup}\label{sec:setup}

\subsection{Fayet-Iliopoulos gauged supergravity}\label{sec:FYsugra}

We consider five-dimensional $\mathcal{N}=2$ supergravity coupled to an arbitrary number $n_V$ of vector multiplets and with a Fayet-Iliopoulos gauging of the R-symmetry~\cite{Gunaydin:1984ak}. We will mostly use the notation of \cite{Gutowski:2004yv}.

The bosonic fields in the theory are the metric $g_{\mu\nu}$, $n_V+1$ Abelian gauge fields $A_\mu^I$, $I=1,\ldots,n_V+1$ (one being the graviphoton in the gravity multiplet), and $n_V$ real scalar fields. The latter are parametrized in terms of $n_V+1$ real functions $X^I$, subject to the constraint
\begin{equation}
\label{constraint}
\frac{1}{6}\, C_{IJK} X^I X^J X^K = 1\,,
\end{equation}
where $C_{IJK}$ is a constant, symmetric tensor. 
The bosonic action in $(-, +,+,+,+)$ signature is:
\begin{equation}
\label{Bulk_action}
S = \frac{1}{2\kappa^2}\!\! \int\! \left[ \left(R - 2  \CV \right)\star \!1 - Q_{IJ} F^I \wedge \star F^J - Q_{IJ}  \diff  X^I \wedge \star\, \diff X^J - \tfrac{1}{6} C_{IJK} A^I \wedge F^J \wedge F^K \right] ,
\end{equation}
where $F^I = \diff A^I$ and $\kappa^2$ is the five-dimensional gravitational coupling constant.

We shall assume that the scalar target space is symmetric. In this case the $C_{IJK}$ tensor satisfies the identity \cite{Gunaydin:1983bi}:
\be\label{3Cinto1C}
C_{IJK} C_{J'(LM} C_{PQ)K'} \,\delta^{JJ'}\delta^{KK'}   \,=\, \frac{4}{3}\,\delta_{I(L} C_{MPQ)}\,.
\ee
We also introduce the lower-index scalars
\begin{equation}\label{XIdown_from_XIup}
X_I=\frac{1}{6} C_{IJK} X^J X^K\ ,
\end{equation}
so that \eqref{constraint} reads
\begin{equation}
\label{constraint_simpler}
X_I X^I =1
\end{equation}
and \eqref{3Cinto1C} implies
\begin{equation}\label{1XupFrom2Xdown}
X^I = \frac{9}{2} C^{IJK} X_J X_K\ ,
\end{equation}
where we defined
\be
C^{IJK}= \delta^{II'}\delta^{JJ'}\delta^{KK'}C_{I'J'K'}\ .
\ee
Note that we also have
\be\label{CuuuXdXdXd}
C^{IJK}X_IX_JX_K = \frac{2}{9}\ .
\ee

The kinetic matrix $Q_{IJ}$ appearing in the action and its inverse $Q^{IJ}$ read:
\begin{align}
Q_{IJ} &= \frac{9}{2}X_IX_J - \frac{1}{2}C_{IJK}X^K\ ,\label{Qmatrix}\\[1mm]
Q^{IJ}&= 2X^IX^J - 6\,C^{IJK}X_K\ ,\label{QmatrixInverse}
\end{align}
and it holds that
\be\label{QXisX}
Q_{IJ}X^J = \frac{3}{2}X_I\ .
\ee

The ungauged supergravity theory has an $\SU(2)$ R-symmetry which rotates the fermion fields. Choosing $n_V+1$ Fayet-Iliopoulos parameters $V_I$, one can gauge a $\U(1)$ subgroup of the R-symmetry by means of the vector field $V_I A^I$. In the bosonic sector all fields remain uncharged and the only consequence of this gauging is to introduce the scalar potential:
\begin{equation}
\mathcal{V} =  - 27\, C^{IJK}V_IV_JX_K\ ,
\end{equation}
as required by supersymmetry.

The Einstein and Maxwell equations following from the action above are: 
\begin{equation}
\label{Einstein_equations}
R_{\mu \nu} - Q_{IJ} F^I_{\mu \kappa} F^{J}{}_{\!\nu}{}^\kappa - Q_{IJ} \partial_\mu X^I \partial_\nu X^J + \frac{1}{6} \,g_{\mu \nu} \left( Q_{IJ}  F^I_{\kappa\lambda}  F^{J \, \kappa \lambda} -4  \CV \right) = 0\ ,
\end{equation}
\begin{equation}
\label{Maxwell_general}
\diff \left( Q_{IJ} \star F^J \right) + \frac{1}{4} C_{IJK} F^J \wedge F^K = 0\ ,
\end{equation}
while the expression of the scalar field equations, that we will not use explicitly, can be found in \cite{Gutowski:2004yv}.

The theory admits a supersymmetric AdS$_5$ vacuum of radius $\ell$,\footnote{Provided $C^{IJK}V_IV_JV_K>0$, which we shall assume.} where the constant values $X^I = \bar X^I$ of the scalars are determined by the Fayet-Iliopoulos parameters as
\begin{equation}
\label{vacuumX}
\bar{X}_I = \ell\, V_I\ .
\end{equation}
When studying the supersymmetry conditions in the next sections, we will find it convenient to use the $\bar{X}_I$ instead of the $V_I$, being understood that these are related as in \eqref{vacuumX}. In terms of such variables the scalar potential may be written as:
\begin{equation}
\label{scalarpot}
\mathcal{V} =  - 6\ell^{-2}\, \bar X^I X_I\ .
\end{equation}

AlAdS solutions to five-dimensional Fayet-Iliopoulos gauged supergravity are expected to describe holographically a dual four-dimensional, $\mathcal{N}=1$ SCFT, possibly deformed by non-trivial background fields or in states different from the conformal vacuum. The supergravity multiplet is dual to the $\mathcal{N}=1$ energy-momentum tensor multiplet (which includes an Abelian R-current), while the supergravity vector multiplets are dual to $\mathcal{N}=1$ Abelian flavour current multiplets. Therefore the field theory deformations that can be studied holographically in this setup are those involving sources or expectation values for the operators in the energy-momentum tensor multiplet and in flavour current multiplets.

Of course, the holographic interpretation is well under control only when the supergravity theory can be uplifted to string theory or M-theory. One such instance is provided by a consistent truncation of type IIB supergravity on $S^5$ \cite{Cvetic:1999xp}, whose SCFT dual is the $\mathcal{N}=4$, $\SU(N)$ super Yang-Mills theory at large $N$. In this case one obtains a Fayet-Iliopoulos gauged supergravity with $n_V = 2$ and with the non-vanishing components of the tensor $C_{IJK}$ being given by $C_{123}=1$, together with those obtained by permutation of the indices. Then the constraint on the scalar fields reads $X^1X^2X^3=1$ and the kinetic matrix is $Q_{IJ}= \frac{9}{2}\,{\rm diag}\left( (X_1)^2,\,(X_2)^2,\,(X_3)^2 \right)$. The scalars in the supersymmetric AdS$_5$ vacuum can be taken as $\bar X^I =1$ for all $I=1,2,3$, which implies $\bar{X}_I = \frac{1}{3}$.\footnote{The bosonic sector of this five-dimensional theory also arises as a consistent truncation of eleven-dimensional supergravity on a space with a boundary \cite{Colgain:2014pha}.}
This consistent truncation retains the vector fields gauging the $\U(1)^3$ Cartan subgroup of the $\SO(6)$ isometries of $S^5$, hence the dual currents are those for the corresponding Cartan subgroup of the $\SO(6)$ R-symmetry in the $\mathcal{N}=4$ super Yang-Mills theory.

\subsection{Supersymmetric solutions with local $\SU(2)\times \U(1)\times \U(1)$ symmetry}\label{sec:susyeqs_general}

We are interested in bosonic, supersymmetric solutions with a local $\SU(2)\times\U(1)\times\U(1)$ symmetry. The existence of a Killing vector is a consequence of supersymmetry, and the form of the solutions depends on whether this is timelike or null \cite{Gutowski:2004yv}. In this paper we just consider the timelike case. The additional $\SU(2)\times\U(1)$ symmetry implies that the supersymmetry conditions reduce to ODE's.
The necessary and sufficient conditions for solutions of this type were given in~\cite{Gutowski:2004yv}. Here we provide a brief summary and then proceed to partially solve such conditions after imposing a simplifying ansatz. In this way we will be left with just two ODE's, generalizing the single ODE obtained in \cite{Gutowski:2004ez} for the minimal gauged supergravity theory. The reader not interested in the derivation can skip to the summary given in Section~\ref{SummarySusyEq}.

A field configuration with the desired symmetry is described by coordinates $y,\rho,\theta,\phi,\hat\psi$ and $\SU(2)$ left-invariant one-forms
\begin{align}\label{def_SU2oneforms_hatted}
\hat\sigma_1 \,&=\, \cos{\hat\psi}\, \diff \theta + \sin{\hat\psi} \sin\theta \,\diff \phi\ , \nn\\
\hat\sigma_2 \,&=\, -\sin{\hat\psi} \,\diff \theta + \cos{\hat\psi} \sin\theta \,\diff \phi \ , \nn\\
\hat\sigma_3 \,&=\,  \diff {\hat\psi} + \cos\theta \,\diff  \phi \  ,
\end{align}
satisfying $\diff \hat\sigma_1 = - \hat\sigma_2 \wedge \hat\sigma_3$, $\diff \hat\sigma_2 = - \hat\sigma_3 \wedge \hat\sigma_1$, $\diff \hat\sigma_3 = - \hat\sigma_1 \wedge \hat\sigma_2$. The hat symbol on $\hat\psi$ (and thus on the $\sigma$'s) distinguishes this coordinate from a different coordinate $\psi$, to be introduced later. The timelike Killing vector determined by supersymmetry will be $V= \frac{\partial}{\partial y}$,
while the other Abelian symmetry will be generated by the left-invariant vector $\frac{\partial}{\partial\hat\psi}$.
The five-dimensional metric takes the form
\begin{equation}
	\label{metric}
	\diff s^2= -f^2(\diff y + w\, \hat\sigma_3)^2 + f^{-1} \left[\,\diff \rho^2 + a^2 (\hat\sigma^2_1+\hat\sigma^2_2) + (2 a a^\prime)^2 \,\hat\sigma^2_3\, \right]\, ,
\end{equation} 
where $a, w, f$ are functions of the radial coordinate $\rho$, and throughout the paper a prime denotes differentiation with respect to $\rho$. The part in square brackets is a K\"ahler metric on a four-dimensional base space $\mathcal{B}$, as required by supersymmetry.\footnote{In general there is an obstruction for a K\"ahler metric to provide a supersymmetric solution~\cite{Figueras:2006xx,Cassani:2015upa}, however this is automatically solved by the particularly symmetric ansatz \eqref{metric}.}

The scalar fields depend on the $\rho$ coordinate only,
$X^I=X^I(\rho)\ ,$
while vector fields contain additional functions $U^I(\rho)$ and are given by:
\begin{equation}
\label{gauge_field}
A^I=X^I f \left(\diff y + w \,\hat\sigma_3\right) + U^I \hat\sigma_3\ ,
\end{equation}
so their field strengths are:
\begin{equation}
\label{Field_Strength}
F^I = - \left(f \, X^I \right)^\prime (\diff y+w\hat\sigma_3)\wedge\diff\rho + \left(f  w^\prime  X^I + \big(U^I\big)' \right) \diff\rho\wedge\hat\sigma_3 -  \left(f  w  X^I + U^I \right) \hat\sigma_1\wedge\hat\sigma_2\ .
\end{equation}
For later use we also record the expression of their Hodge dual (we inherit from \cite{Gutowski:2004yv} the choice of $\diff y\wedge\diff\rho\wedge\hat\sigma_1\wedge\hat\sigma_2\wedge\hat\sigma_3$ for the positive orientation): 
\begin{align}
\label{Field_strength_star}
\star F^I &= 2a^3a'f^{-2} \left(f X^I \right)^\prime \hat\sigma_{123} + \frac{af}{2a^\prime} \left(f w^\prime X^I + (U^I)' \right) (\diff y+w\hat\sigma_3)\wedge\hat\sigma_1\wedge\hat\sigma_2 \nn\\[1mm]
&\quad \, - \frac{2a'}{a}f \left(f  w  X^I + U^I \right) \diff y \wedge\diff\rho\wedge \hat{\sigma}_3 \ .
\end{align}

The functions $a(\rho), w(\rho),  f(\rho), X^I(\rho), U^I(\rho)$ controlling the solution are determined by the following equations \cite{Gutowski:2004yv}:
\begin{equation}
\label{eqforf}
f=  f_{\rm min} \,\bar{X}^I X_I\ ,
\end{equation}
\begin{equation}
\label{eqforU}
\left(a^2 U^I\right)' = 36\frac{\epsilon}{\ell}\,   a^3 a^\prime f^{-1}\, C^{IJK} \bar X_J X_K\ ,
\end{equation}
\begin{equation}
\label{eqforw}
f^{-1} X_I \left(a^{-2} U^I\right)' = - \frac{2}{3} \left(a^{-2} w\right)'\ ,
\end{equation}
\begin{equation}
\label{eqfora}
\bar{X}_I U^I = \frac{\epsilon \ell}{3}\, p\ ,
\end{equation}
\begin{equation}
\label{maxwelleq}
\left[a^3 a^\prime \left(f^{-1} X_I\right)' + \frac{\epsilon}{\ell}\bar{X}_I a^2 w  + \frac{1}{12} C_{IJK} U^J U^K \right]' = 0 \,.
\end{equation}
These are obtained by combining the supersymmetry conditions and the Maxwell equation. In particular, \eqref{maxwelleq} follows from the Maxwell equation.
Here $\epsilon=\pm 1$ is an arbitrary sign choice related to the versus of rotation of the solution along $\frac{\partial}{\partial\hat\psi}$.
The function
\begin{equation}
\label{eqforfminimal}
f_{\rm min} = \frac{12 \, a^2  a^\prime}{\ell^2 (a^2 a^{\prime\prime\prime} - a^\prime + 7 a a^\prime a^{\prime \prime} + 4(a^\prime)^3)} \,
\end{equation}
is the expression for $f$ that is obtained when working in minimal gauged supergravity~\cite{Gutowski:2004ez}. It has a geometric meaning as it is proportional to the inverse scalar curvature $R_\mathcal{B}$ of the four-dimensional K\"ahler base $\mathcal{B}$,  $f_{\rm min} = -\frac{24}{\ell^2 R_\mathcal{B}}$.
Moreover in \eqref{eqfora} we have introduced the function:
\begin{equation}
\label{P}
p = -1 + 2 a a^{\prime\prime} + 4 (a^\prime)^2 \,.
\end{equation}
It will be useful to note the identity\footnote{The function $p$ determines the Ricci form on the K\"ahler base as $\mathcal{R}=\epsilon\,\diff(p\, \hat\sigma_3)$. The identity \eqref{curvature_from_ricci_pot} expresses the fact that in K\"ahler geometry the trace of the Ricci form is proportional to the Ricci scalar, $J^{mn}\mathcal{R}_{mn}=R$. Here $J = -\epsilon\, \diff(a^2\hat\sigma_3)$ is the K\"ahler form on the K\"ahler base $\mathcal{B}$ \cite{Gutowski:2004yv}.
}
\be\label{curvature_from_ricci_pot}
a^3a'f_{\rm min}^{-1} = \frac{\ell^2}{24}\left( a^2p\right)'\,. 
\ee

We now proceed to manipulate the equations of \cite{Gutowski:2004yv} given above and partially solve them.
With no loss of generality, we can express the vector of functions $X_I(\rho)$ by separating the component along the constant vector $\bar{X}_I$ and the orthogonal ones:
\be\label{scalaransatz}
f^{-1}X_I = f_{\rm min}^{-1}\bar X_I + h_I\ ,
\ee
where $h_I$ are functions of $\rho$ satisfying
\be\label{barXhis0}
\bar{X}^Ih_I = 0 \ ,
\ee
while the component along $\bar{X}_I$ has already been fixed using~\eqref{eqforf}.
Plugging \eqref{scalaransatz} in the constraint \eqref{CuuuXdXdXd}, we find that $f$ is expressed as:
\be\label{f_from_a}
f = \left(f_{\rm min}^{-3} + \frac{27}{2}f_{\rm min}^{-1}C^{IJK}\bar{X}_I h_Jh_K + \frac{9}{2}C^{IJK}h_Ih_Jh_K  \right)^{-1/3}.
\ee
Recalling the identity \eqref{curvature_from_ricci_pot}, equation \eqref{eqforU} for $U^I$ becomes
\begin{equation}
\left(a^2 U^I\right)' = \frac{\epsilon\ell}{3} \bar{X}^I \left(a^2p\right)'  + \frac{36\epsilon}{\ell} a^3a' \, C^{IJK}\bar{X}_J h_K \,.
\end{equation}
It is convenient to trade $h_I$ for some new functions, $H_I(\rho)$, defined as
\be\label{h_from_H}
h_I = \frac{H_I'}{a^3a'}\,.
\ee
In this way the equation for $U^I$ can be solved as
\be
\label{UI_from_a}
U^I = \frac{\epsilon\ell}{3}\bar X^I p +  \frac{36\epsilon}{\ell a^2} C^{IJK}\bar{X}_J H_K + \frac{U_0^I}{a^2}\ ,
\ee
where $U_0^I$ are integration constants. Compatibility of this solution with \eqref{eqfora} implies $\frac{8\epsilon}{\ell}\bar{X}^I H_I + \bar{X}_I U_0^I=0$. In the following we will choose $U_0^I = 0\,$\footnote{We made a preliminary analysis with $U_0^I \neq 0$ and found no regular solutions due to a divergence appearing in the perturbative expansion at small $\rho$.} and thus require that
\be\label{XHis0}
\bar{X}^IH_I = 0\ .
\ee

So far we have expressed $X_I, f$ and $U^I$ in terms of $a$ and $H_I$. Next we use these findings to manipulate eq.~\eqref{eqforw} containing $w$ and the Maxwell equation~\eqref{maxwelleq}, following a strategy used in Section 4 of \cite{Gutowski:2004ez} in the context of minimal gauged supergravity.
Introducing
\be\label{expr_g}
g = -\frac{a'''}{a'}-3 \frac{a''}{a}-\frac{1}{a^2} + 4 \frac{a'^2}{a^2}\ ,
\ee
we notice that
\be
\left(a^{-2} p\right)' = -\frac{2a'g}{a}\ .
\ee
Then eq.~\eqref{eqforw} becomes
\be\label{eqforw_ter}
\frac{a}{2a'}(a^{-2}w)' \,\equiv\,\frac{w'}{2aa'} - \frac{w}{a^2} \,=\, \epsilon\left[ \frac{\ell}{2}f_{\rm min}^{-1}\,g - \frac{27a}{\ell\, a'}\,\bar{X}_IC^{IJK}\frac{H_J'}{a^3a'}\left(\frac{H_K}{a^4} \right)^{\!\!'}\;\right] \ .
\ee

We now massage the Maxwell equation \eqref{maxwelleq}. After some computations involving the identity \eqref{3Cinto1C}, we find that
\be
C_{IJK}U^JU^K = \frac{2\ell^2}{3}\bar{X}_I\, p^2 + \frac{8p}{a^2}H_I + \frac{288}{\ell^2a^4} \bar{Q}_{IJ} (CHH)^J\ ,
\ee
where we used the shorthand notation $(CHH)^J=C^{JKL}H_KH_L$,
while by $\bar{Q}^{IJ}$ we denote the kinetic matrix~\eqref{QmatrixInverse} evaluated on $X=\bar X$.
Eq.~\eqref{maxwelleq} then becomes
\be\label{MaxwellEqFull}
\left[ a^3 a^\prime \left( f_{\rm min}^{-1}\bar{X}_I + \frac{H'_I}{a^3a'} \right)' + \bar{X}_I\left(\frac{\epsilon}{\ell} a^2 w  + \frac{\ell^2p^2}{18}\right) + \frac{2p}{3a^2} H_I + \frac{24}{\ell^2a^4}\bar{Q}_{IJ} (CHH)^J  \right]' = 0 \,.
\ee
The component along $\bar{X}_I$, which is obtained by contracting with $\bar{X}^I$, reads
\be\label{MaxwellEqParallel}
\left[ a^3 a^\prime \left( f_{\rm min}^{-1} \right)' + \frac{\epsilon}{\ell} a^2 w  + \frac{\ell^2p^2}{18} + \frac{36}{\ell^2a^4}  C^{IJK}\bar{X}_{I} H_JH_K  \right]' = 0 \,.
\ee
The components having vanishing contraction with $\bar{X}^I$, which are given by ${\rm Maxw}_I - \bar{X}_{I} \bar{X}^J{\rm Maxw}_J$, where 
${\rm Maxw}_I$ is eq.~\eqref{MaxwellEqFull}, read instead
\be\label{Maxwell_orthog}
\left[ H_I'' - \left(\frac{3a'}{a}+\frac{a''}{a'}\right) H_I' + \frac{2p}{3a^2}H_I + \frac{24}{\ell^2a^4}\left( \bar{Q}_{IJ} - \frac{3}{2}\bar{X}_I \bar{X}_J \right) (CHH)^J   \right]'=0\,.
\ee
Eq.~\eqref{MaxwellEqParallel} can also be written as
\be\label{MaxwellEqParallel_bis}
\frac{w'}{2aa'} + \frac{w}{a^2} =- \frac{\epsilon\ell}{2}\left[ \nabla^2 (f_{\rm min}^{-1})  + 8\ell^{-2}f_{\rm min}^{-2} -\frac{\ell^2g^2}{18} + \frac{36}{\ell^2a^3a'}\bar{X}_IC^{IJK}\left(\frac{H_JH_K}{a^4}\right)'\,  \right] \,,
\ee
where
\begin{equation}
\nabla^2 f_\text{min}^{-1} = \frac{1}{a^3 a'}\left( a^3 a' \left(f_\text{min}^{-1}\right)'\right)'
\end{equation}
is the Laplacian of $f_\text{min}^{-1}$ on the K\"ahler base $\mathcal{B}$.

Combining \eqref{MaxwellEqParallel_bis} with \eqref{eqforw_ter} one can eliminate $w'$ and solve for $w$ as
\begin{align}
\label{solw_runningX}
w \,&=\, -\frac{\epsilon \ell a^2}{4}\left\{ \nabla^2 (f_{\rm min}^{-1}) + \frac{8}{\ell^2}f_{\rm min}^{-2}  - \frac{\ell^2g^2}{18} + f_{\rm min}^{-1}\,g  \right.\nn \\[1mm]
 &\qquad\qquad\quad + \left.\frac{36}{\ell^2a^3a'}\bar{X}_IC^{IJK}\left[\left( \frac{H_JH_K}{a^4}\right)' - \frac{3a}{2a'} H_J' \left( \frac{H_K}{a^{4}}\right)'\, \right]\right\}\,.
\end{align}
Plugging this back into either \eqref{eqforw_ter} or \eqref{MaxwellEqParallel}, we finally arrive at
\begin{align}
\label{eqfora_runningX}
&\bigg( \nabla^2 f_\text{min}^{-1} + \frac{8}{\ell^2} f_\text{min}^{-2} - \frac{\ell^2 g^2}{18} + f_\text{min}^{-1}\, g \bigg)' + \frac{4 a' g}{a f_\text{min}}\nn\\[1mm]
+& \bar{X}_IC^{IJK}\!\left\{\frac{36}{\ell^2a^3a'}\left[\left(\! \frac{H_JH_K}{a^4}\!\right)' - \frac{3a}{2a'} H_J'\left(\!\frac{H_K}{a^4}\!\right)'\; \right] \right\}' -\frac{216}{\ell^2}\bar{X}_IC^{IJK} \frac{H_J'}{a^3a'}\!\left(\!\frac{H_K}{a^4}\!\right)'  = 0\,.
\end{align}

We have thus partially solved the system of equations \eqref{eqforf}--\eqref{maxwelleq} for $a$, $f$, $w$, $U^I$, $X_I$, and are left with the equations~\eqref{Maxwell_orthog}, \eqref{eqfora_runningX} involving just the unknown functions $H_I$ and $a$. Eq.~\eqref{Maxwell_orthog} is third order in the variable $\rho$, while eq.~\eqref{eqfora_runningX} contains up to six derivatives.

When $H_I=0$, the equations above simplify considerably and reduce to the supersymmetry conditions obtained in minimal gauged supergravity  \cite{Gutowski:2004ez}. Indeed \eqref{f_from_a} yields $f = f_{\rm min}$ while from \eqref{scalaransatz} we see that the scalars are set to the constant value taken in the AdS$_5$ solution, $X^I = \bar{X}^I$. The expression \eqref{gauge_field} for the gauge fields becomes
\be\label{A_minimal}
A^I= \bar{X}^I A\,, \quad {\rm with}\quad A = f \left(\diff y + w \,\hat\sigma_3\right) + \frac{\epsilon\ell}{3} p\, \hat\sigma_3 \ 
\ee
being the graviphoton of minimal gauged supergravity.
Moreover, \eqref{Maxwell_orthog} trivializes while eqs.~\eqref{eqforw_ter}, \eqref{MaxwellEqParallel}, \eqref{solw_runningX}, \eqref{eqfora_runningX} reduce to those of the minimal case given in~\cite{Gutowski:2004ez}. We thus conclude that our equations~\eqref{Maxwell_orthog}, \eqref{eqfora_runningX} provide a direct generalization of the minimal supersymmetry equation of~\cite{Gutowski:2004ez} to the case with an arbitrary number of vector multiplets, where both the gauge and the scalar fields are running.

\subsection{A simplifying ansatz}\label{sec:simpl_ansatz}

So far we have manipulated the original supersymmetry equations of \cite{Gutowski:2004yv} without any restriction,\footnote{Apart for fixing the integration constants $U_0^I=0$ when solving for $U^I$.} arriving at eqs.~\eqref{Maxwell_orthog}, \eqref{eqfora_runningX}. 
We now impose the ansatz
\be\label{ansatz_H}
H_I = q_I H\,,\qquad I = 0,\ldots,n_V\ ,
\ee
where $H(\rho)$ is a real function and $q_I$ is a constant vector, which for consistency with \eqref{XHis0} must be orthogonal to $\bar{X}_I$,
\be\label{barXq=0}
\bar{X}^I q_I = 0\,.
\ee 
Although this ansatz will not be enough for solving the equations analytically, it will be helpful while performing the perturbative and numerical analysis in the next sections.

Plugging our ansatz in, 
eq.~\eqref{Maxwell_orthog} becomes
\be\label{Maxwell_orthog_Bis}
q_I\left[ H'' - \left(\frac{3a'}{a}+\frac{a''}{a'}\right) H' + \frac{2p}{3a^2}H \right]' - \frac{4}{\ell^2}\,W_I\left( \frac{H^2}{a^4}   \right)'=0\ ,
\ee
where the constant vector $W_I$ is defined as
\be
\label{W_parallel}
W_I = \left( -6\bar{Q}_{IJ} + 9\bar{X}_I \bar{X}_J \right) C^{JKL}q_Kq_L \ .
\ee

If $W_I = 0$, then one can see that necessarily $q_I=0$,\footnote{Indeed multiplying $\left( -6\bar{Q}_{IJ} + 9\bar{X}_I \bar{X}_J \right)  (Cqq)^J=0$ by $\bar Q^{-1}$ and using \eqref{QXisX} we obtain
$(Cqq)^I = (C\bar{X}qq) \bar{X}^I$. 
 Contracting \eqref{3Cinto1C} with four $q$'s one finds that this implies $(C\bar{X}qq)=0$. This in turn means that $q_I = 0$, see appendix~\ref{app:conditions_qI} for details. 
\label{ftnote:proofqis0}} that is $H_I=0$. As discussed at the end of Section~\ref{sec:susyeqs_general}, in this case there would be no running scalars and we would be left with the equations of minimal gauged supergravity.
Therefore we assume $W_I \neq 0$. We should now distinguish whether the constant vectors $q_I$ and $W_I$ are linearly dependent or not. If they are independent, then their coefficients in \eqref{Maxwell_orthog_Bis}  have to vanish separately. In this case, from the term proportional to $W_I$ we obtain 
\be\label{GR2sol_H}
H = {\rm const}\, a^2\ ,
\ee 
while the rest of~\eqref{Maxwell_orthog_Bis} has, up to trivial symmetries involving shifts and rescalings of the coordinate $\rho$, the general solution:
\be\label{GR2sol_a}
a \,=\, \alpha\ell \sinh (\rho/\ell) \,,
\ee
where $\alpha$ is a parameter. This also satisfies \eqref{eqfora_runningX} and is just the solution found in~\cite{Gutowski:2004yv}, leading to an asymptotically AdS black hole whose boundary is conformally flat.
  
Therefore new solutions within the ansatz \eqref{ansatz_H} may only be found if we assume that the vectors $W_I$ and $q_I$ are parallel to each other. Since the overall scale of $q_I$ is immaterial (as it can always be reabsorbed in the function $H$), there is no loss of generality in assuming $W_I = q_I$. That is, we take
\be\label{cond_on_q}
\left( -6\bar{Q}_{IJ} + 9\bar{X}_I \bar{X}_J \right) C^{JKL}q_Kq_L \,=\,  q_I\,.
\ee
Note that this implies~\eqref{barXq=0}. Thus we have a system of $n_V+1$ equations for $n_V+1$ unknowns $q_I$,  which in general determines  the $q_I$.
In appendix~\ref{app:conditions_qI} we show that \eqref{cond_on_q} also implies
\be\label{CqqMainText}
C^{IJK}q_Jq_K = -\frac{1}{18} \bar{X}^I + \bar{Y}^I\ , \quad\text{where}\quad
\bar{Y}^I = 
C^{IJK} \bar{X}_J q_K \ ,
\ee
\be
C^{IJK}\bar{X}_Iq_Jq_K =  C^{IJK}q_Iq_Jq_K = - \frac{1}{18}\ .\label{CXqq_and_Cqqq_MainText}\\[1mm]
\ee
These relations are enough for simplifying the supersymmetry conditions of Section~\ref{sec:susyeqs_general} (with ansatz \eqref{ansatz_H} plugged in) in such a way that one can look for solutions independently of the specific values taken by the $q_I$. The resulting equations are collected below.

\subsection{Summary of supersymmetry equations}\label{SummarySusyEq}

We summarize here the result of using the ansatz \eqref{ansatz_H} into the conditions for a timelike supersymmetric solution to Fayet-Iliopoulos gauged supergravity with local $\SU(2)\times\U(1)\times\U(1)$ symmetry, discussed in Section \ref{sec:susyeqs_general}. We have found that a solution is obtained by solving the following coupled ODE's for the functions $a(\rho)$, $H(\rho)$:
\be\label{eqforH}
\left[\, H'' - \left(\frac{3a'}{a}+\frac{a''}{a'}\right) H' + \frac{2p}{3a^2}H  -\frac{4}{\ell^2} \frac{H^2}{a^4}  \: \right]' = \,0\ ,
\ee
\begin{align}\label{new_eqfora}
&\bigg( \nabla^2 f_\text{min}^{-1} + \frac{8}{\ell^2} f_\text{min}^{-2} - \frac{\ell^2 g^2}{18} + f_\text{min}^{-1}\, g \bigg)' + \frac{4 a' g}{a f_\text{min}}\nn\\[2mm]
-&\, \left\{\frac{2}{\ell^2a^3a'}\left[\left( \frac{H^2}{a^4}\right)' - \frac{3a}{2a'} H'\left(\frac{H}{a^4}\right)'\; \right] \right\}' +\frac{12}{\ell^2}  \frac{H'}{a^3a'}\left(\frac{H}{a^4}\right)'  = 0\ ,
\end{align}
where we recall that $f_\text{min}$, $p$ and $g$ are the functions of $a$ and its derivatives given in~\eqref{eqforfminimal}, \eqref{P}, and \eqref{expr_g}, respectively.
Once a solution for $a$ and $H$ is obtained, the five-dimensional supergravity fields are fully determined.
The metric and the gauge fields take the form \eqref{metric}, \eqref{gauge_field}, where the 
 functions $f$, $w$ and $U^I$ read:
\be
\label{f_from_a_ansatz}
f = \bigg[\, f_{\rm min}^{-3} - \frac{3}{4}f_{\rm min}^{-1} \bigg(\frac{H'}{a^3a'}\bigg)^2 - \frac{1}{4} \bigg(\frac{H'}{a^3a'}\bigg)^3\; \bigg]^{-1/3},
\ee
\be
\label{w_from_a_ansatz}
w = -\frac{\epsilon \ell a^2}{4}\left\{ \nabla^2 (f_{\rm min}^{-1}) + \frac{8}{\ell^2}f_{\rm min}^{-2}  - \frac{\ell^2g^2}{18} + f_{\rm min}^{-1}\,g - \frac{1}{2 \ell^2 a^3 a^\prime} \bigg[\bigg(\frac{H^2}{a^4}\bigg)^\prime - \frac{3a}{2a^\prime} H \bigg(\frac{H}{a^4}\bigg)^\prime\, \bigg] \right\} ,
\ee
\begin{equation}
\label{UI_from_ansatz}
U^I = \frac{\epsilon\ell}{3}\bar X^I p +  \frac{36\epsilon}{\ell}\, \bar{Y}^I\, \frac{H}{a^2}\ .
\end{equation}
The scalar fields $X^I$ are computed from
\be
\label{scalars_down_from_ansatz}
X_I =  \bar{X}_I f  f^{-1}_\text{min} + q_I  f\frac{H^\prime}{a^3 a^\prime} 
\ee
using \eqref{1XupFrom2Xdown}, and read
\be
\label{scalars_up_from_ansatz}
X^I \,=\,   \bar{X}^I  f^2\bigg[f_{\text{min}}^{-2}  -\frac{1}{4} \bigg(\frac{H^\prime} {a^3 \, a^\prime} \bigg)^{\!2}\;\bigg] + 9 \, \bar{Y}^I f^2  \left[f_{\text{min}}^{-1} + \frac{H^\prime} {2a^3 \, a^\prime} \right]\frac{H^\prime} {a^3 \, a^\prime}  \ .
\ee
Note that they split into a part aligned to $\bar{X}^I$ and one aligned to $\bar{Y}^I$.
We recall that the constants $\bar X^I$ are the values of the scalar fields in the AdS$_5$ vacuum, while $\bar{Y}^I = C^{IJK} \bar{X}_J q_K$, with the constants $q_K$ being in general determined by condition~\eqref{cond_on_q}.
For instance, for the $\U(1)^3$ theory that is obtained as a consistent truncation of type IIB supergravity on $S^5$ described at the end of Section~\ref{sec:FYsugra}, it is easy to see that the only allowed choices for the $q_I$ are either
$q_1 = q_2 = \frac{1}{6}$, $q_3 = -\frac{1}{3}$,
or the similar expressions obtained by cyclically permuting the indices $1,2,3$. This implies that the $\bar Y^I$ take the values $\bar Y^1 = \bar Y^2 = -\frac{1}{18}$, $\bar Y^3 = \frac{1}{9}$ (or their cyclic permutations).

These solutions generically preserve two supercharges.
When $H=0$ the expressions above reduce to the conditions for supersymmetric solutions to minimal gauged supergravity obtained in~\cite{Gutowski:2004ez}.

%%%%%%%%%%%%%%%%%%%%%%%%%%%%%%%%%%%%%%%%%%%%%%%%%%%%%%%%%%%%%%%%%%%%%%%%%%%%%%%%%%%%%%
\section{The solution}\label{sec:Solution}

In this section we solve perturbatively the equations presented above. For simplicity we will set $\ell = 1$ and make the sign choice $\epsilon = + 1$.

\subsection{Near-boundary solution}\label{sec:near_boundary_sol}

We study our equations \eqref{eqforH} and \eqref{new_eqfora} perturbatively around $\rho \to \infty$, which as we will see corresponds to a limit where a conformal boundary is approached.  We assume the following asymptotic expansions for the unknown functions $a$ and $H$:
\begin{align}
a(\rho) & = a_0 e^\rho \bigg[ 1 + \sum_{k \geq 1} \, \sum_{0 \leq n \leq k} \, a_{2k,n} \, \rho^n \, \left(a_0 \, e^\rho \right)^{-2k} \bigg] \notag \\
& = a_0 e^\rho \left[1 + \left(a_{2,0} + a_{2,1} \rho \right) \frac{e^{-2\rho}}{a^2_0} + \left(a_{4,0} + a_{4,1} \, \rho + a_{4,2} \, \rho^2 \right) \frac{e^{-4\rho}}{a^4_0}  + \dots \right]\ , 
\end{align}
\begin{align}
H(\rho) & = a_0^4 e^{4 \rho} \sum_{k \geq 0} \, \sum_{0 \leq n \leq k} \, H_{2k,n} \, \rho^n \, \left(a_0 \, e^\rho \right)^{-2k}  \notag \\
& = a_0^4 e^{4\rho} \bigg[H_{0,0} + \left(H_{2,0} + H_{2,1} \rho \right) \frac{e^{-2\rho}}{a^2_0} + \left(H_{4,0} + H_{4,1} \, \rho + H_{4,2} \, \rho^2 \right) \frac{e^{-4\rho}}{a^4_0} + \dots \bigg]\ ,
\end{align}
with $a_0\neq 0$. Note that the expansion of $a$ only involves odd powers of $e^\rho$; we could have included terms involving even powers but they would have been set to zero by the equations. For the same reason the expansion of $H$ only involves even powers of $e^\rho$. We solved \eqref{eqforH} and \eqref{new_eqfora} perturbatively up to order $\CO(e^{-10 \rho})$ and found a family of solutions controlled by eight free parameters. Renaming them for convenience, these are: 
\be
a_0\,,\quad a_2\equiv a_{2,0}\,,\quad c \equiv a_{2,1}\,,\quad a_4 \equiv a_{4,0}\,,\quad a_6 \equiv a_{6,0}\ ,\nn
\ee
\be
H_2 \equiv H_{2,0}\,,\quad H_4 \equiv H_{4,0}\,, \quad\HL \equiv H_{2,1}\ .
\ee
 We report here the first terms in the expansion of $H$ and~$a$:
\begin{align}\label{UVsola}
a(\rho) &= a_0 \,e^\rho + \left(a_2 + c \rho \right) \frac{e^{-\rho}}{a_0}  \nn\\
& \!\! +\! \bigg[ a_4 +\frac{2-16 a_2 -5 c }{12} c \rho + \frac{3}{8} \big(2H_2+3\HL\big)\HL\rho - \frac{2}{3} c^2 \rho^2 + \frac{3}{8}\HL^2\rho^2 \bigg] \frac{e^{-3 \rho}}{a_0^3} + \CO(e^{-4 \rho})\,,\\[1mm]
\label{UVsolH}
H(\rho) &= \big(H_2 + \HL \rho \big) a^2_0  e^{2 \rho} + H_4 +  2\big(H_2 +  \HL\big)\HL\rho + \frac{1}{6}\big(4 a_2 \HL + 4 c H_2 - 2 c \HL + \HL\big)\rho \nn\\
&\quad  + \Big(\frac{2}{3}c \HL  + \HL^2\Big)\rho^2 + \CO(e^{-2 \rho})\ .
\end{align}
Notice that the backreaction of the fields in the supergravity vector multiplets introduces a dependence on $H_4$, $H_2$, $\HL$ in the metric functions.\footnote{We also found a different solution for $H(\rho)$, having $H_{0,0} = 1$ (while $H_{0,0}=0$ in \eqref{UVsolH}) and governed by the free parameter $H_{4,0}$. However the leading term of the corresponding metric turns out to be of order $\CO(e^{4 \rho})$, indicating that the latter is not AlAdS. %Indeed an AlAdS metric should diverge at most as $\CO(e^{2 \rho})$ as $\rho \to \infty$. 
 For this reason we will not discuss this other solution in the following.}

Starting from the solution for $a$ and $H$, we can construct the asymptotic form of  the supergravity fields by using the formulae given in the previous section.
 In the following we only provide the leading order terms, while in Appendix \ref{app:FeffermanGraham} we display the relevant subleading terms after turning the solution to Fefferman-Graham form. This also shows that the solution is AlAdS$_5$.

We find it convenient to trade the parameter $c$ for a new parameter
\be
v^2 = 1-4c\ ,
\ee 
which will turn out to control the squashing of the three-sphere. We also change the coordinates $y,\,\hat\psi$ into new coordinates $t,\, \psi$, defined as:
\be
\label{change_psi}
 y = t \ , \qquad\qquad
 \hat{\psi} = \psi - \frac{2}{v^2} \, t \ .
\ee
These lead to a static (rather than stationary) metric on the conformal boundary. In these coordinates, the supersymmetric Killing vector $V$ reads
\begin{equation}
\label{Killing_tpsi}
	V = \frac{\partial}{\partial y} 
	= \frac{\partial}{\partial t} + \frac{2}{v^2} \frac{\partial}{\partial \psi}\ .
\end{equation}
The five-dimensional metric and the gauge fields in the new coordinates take the general form:
\begin{equation}\label{comp_5dmetric}
\diff s^2 = g_{\rho \rho} \diff  \rho^2 + g_{\theta \theta} (\sigma_1^2 + \sigma_2^2) + g_{\psi \psi} \sigma_3^2 + g_{tt} \diff t^2 + 2 g_{t \psi} \, \sigma_3 \, \diff t\ ,
\end{equation}
\begin{equation}\label{comp_5dgaugefield}
A^I = A^I_t \, \diff t + A^I_\psi \, \sigma_3\ ,
\end{equation}
where the one-form $\sigma$'s are defined as the $\hat\sigma$'s in \eqref{def_SU2oneforms_hatted}, but using $\psi$ instead of $\hat\psi$.

We find that at leading order the five-dimensional metric reads:
\be
\diff s^2 = \diff \rho^2 + e^{2\rho}\,\diff s^2_{\rm bdry}+ \ldots\ ,
\ee
where the metric on the conformal boundary is:
\be\label{bdrymetric}
\diff s^2_{\rm bdry} \ = \  (2a_0)^2\left[ - \frac{1}{v^2} \diff t^2 + \frac{1}{4}\left(\sigma_1^{\,2} + \sigma_2^{\,2}  +   v^2\sigma_3^{\,2}\right) \right]\ .
\ee
As anticipated this is static in the chosen coordinates.
The three-dimensional part of the metric involving the $\sigma$'s is locally the metric on a Berger three-sphere, with $v$ controlling the $\SU(2)\times \U(1)$ invariant squashing of the Hopf fiber.
 
 The gauge fields $A^I$ have a part along $\bar{X}^I$ and a part along $\bar{Y}^I$. These  can be isolated by contracting $A^I$ with either $\bar{X}_I$ or
 \be
 \bar{Y}_I \equiv -18\, q_I\ . 
\ee
 Indeed these quantities satisfy the relations 
 \be
 \bar{X}_I \bar{X}^I = \bar{Y}_I \bar{Y}^I = 1\ ,\qquad\bar{X}_I \bar{Y}^I = \bar{Y}_I \bar{X}^I = 0\ .
 \ee 
By doing so we obtain the following expressions at leading order:
\be\label{bdry_AX}
\bar{X}_I A^I = \frac{v^2+2 }{3 \, v^2}\, \diff t + \frac{1}{3}  (v^2-1) \,\sigma_3 + \CO(e^{- \rho})
\ee
 and 
 \be
 \bar{Y}_I A^I = 36\,\frac{\HL}{v^2} \,\diff t - 18  \, \HL \,\sigma_3 + \CO(e^{- \rho})\ .
 \ee
Note that both $\bar{X}_I A^I$ and $\bar{Y}_I A^I$ have a non-trivial boundary field-strength proportional to $\sigma_1\wedge\sigma_2$.

Evaluation of the scalar fields $X^I$ yields:
\begin{equation}
X^I = \bar{X}^I + 9 \, \bar{Y}^I \left(2 H_2 + \HL +2 \HL \, \rho \right)\frac{e^{-2\rho}}{a_0^2} + \CO(e^{-3\rho})\ .
\end{equation}

In the AdS$_5$ solution, our scalar fields have mass $m^2\ell^2 = -4$, hence the conformal dimension of the dual operator, following from the well-known formula $m^2\ell^2 = \Delta(\Delta-4)$, is $\Delta = 2$. This is also reflected in the asymptotic behavior displayed above. 

Inspection of the solution in Fefferman-Graham coordinates (see Appendix \ref{app:FeffermanGraham}) shows that the free parameters $a_0,c$ and $\HL$ specify the boundary conditions of the bulk fields and are thus associated to sources in the dual field theory.
As already apparent from the expressions above, $a_0$ and $c$ determine both the metric and the value of $\bar{X}_I A^I$ at the conformal boundary, while $\HL$ fixes the asymptotic mode of the scalar fields. The three parameters $a_0$, $c$ and $\HL$ together also determine $\bar{Y}_IA^I$. 
The remaining parameters $a_2$, $a_4$, $a_6$, $H_2$, $H_4$ instead control dual field theory one-point functions. In particular, $H_2$ controls the normalizable mode of the scalar fields.
We will come back to the holographic interpretation of our solution in Section~\ref{sec:properties_sol}.

\subsection{Near-horizon solution}\label{sec:near_horizon_sol}

We now turn to solving eqs.~\eqref{eqforH}, \eqref{new_eqfora} near to $\rho =0$.  
 We assume that both the $a$ and $H$ functions can be Taylor expanded as:
\begin{align}
a(\rho) \,&=\, \air_0 + \air_1 \, \rho + \air_2 \, \rho^2 + \dots\ , \nn \\[1mm]
H(\rho) \,&=\, \hir_0 + \hir_1 \, \rho + \hir_2 \, \rho^2 + \dots\ .
\end{align}
We are interested in solutions that either close off regularly or meet an event horizon when $\rho\to 0$. In both cases, given the form \eqref{metric} of the metric we should take $\air_0 = 0$ in the expansion above. Furthermore the form of the supersymmetry equations allows us to assume $\air_1 > 0$ with no loss of generality (we are not interested in solutions with $\air_1 = 0$).

We solved equations~\eqref{eqforH}, \eqref{new_eqfora} order by order in powers of $\rho$, up to $\mathcal{O}(\rho^{18})$. 
We found different branches of solutions, most of them corresponding to the small-$\rho$ expansion of \eqref{GR2sol_H}, \eqref{GR2sol_a}, that is the well-known solution of \cite{Gutowski:2004yv}. However we also obtain one interesting branch of solutions to~\eqref{eqforH}, yielding the following expression for $H$:
\begin{align}
& H (\rho) \,=\, \hir\, \air^2 \rho^2 + \frac{2 \, \air  \, \air_2 \, \hir \, (2 - 3 \, \air^2 + 24 \, \hir)}{
2 + \air^2 + 24 \hir}\, \rho^3  \notag \\[1mm]
&\, + \frac{\hir}{81}  \bigg[ 81(\air_2^2 + 2 \air \air_3) - \frac{
16 (-2 + 17 \air^2) \air_2^2}{1 - 4 \air^2 + 12 \hir} - \frac{
288 \air^2 \air_2^2\, (2 + \air^2)}{(2 + \air^2 + 24 \hir)^2} - \frac{
8 (8 + 175 \air^2) \air_2^2}{2 + \air^2 + 24 \hir} \bigg] \rho^4 \nn \\[1mm]
&\, + \CO(\rho^5)\ ,
\end{align}
where we  defined 
\be
\air\equiv\air_1\ ,\qquad \hir \equiv \frac{\hir_2}{\air_1^2}\ ,\ee
  We see that  $H(\rho)$ is entirely determined by $\eta$ and the coefficients of $a(\rho)$. These in turn are controlled by eq.~\eqref{new_eqfora}. Analysis of the latter requires distinguishing different cases, as we now describe. 
The first non-trivial order of~\eqref{new_eqfora} yields: 
\be\label{first_option_IR}
\air_2 \left( 8 +13\air^2 + \frac{576\, \hir^2}{2+\air^2+ 24\,\hir}  \right)\,=\,0\ ,
\ee
so we have to set either $\air_2$ or the parenthesis to zero. In this paper we will choose $\air_2=0$ and will not discuss the other option. One reason is that this is also the condition that is imposed when working in minimal gauged supergravity  \cite{Cassani:2014zwa}, and we would like our solutions  to admit a limit such that they are contained in the latter.\footnote{In the  analysis of \cite{Cassani:2014zwa}, the condition corresponding to \eqref{first_option_IR} was $\air_2\left( 8 +13\air^2  \right)=0$, hence the choice $\air_2=0$ was the only possible.}
At the next order we get:
\be\label{second_option_IR}
\air_4 \left( -8 + 11 \air^2 - \frac{576 \,\hir^2}{2 -23\air^2 +24 \hir} \right) \,=\, 0\,.
\ee
When $\eta=0$ this reduces to $\air_4 \left( -8 + 11 \air^2\right) \,=\, 0$, that is
 the equation found in \cite{Cassani:2014zwa} for the minimal theory. In \cite{Cassani:2014zwa}, the choice $\air_4=0$ led to either the solution of \cite{Gutowski:2004ez} (given by \eqref{GR2sol_a} above) or to a regular soliton that was identified as the gravity dual of the vacuum state of four-dimensional superconformal field theories on a squashed $S^3\times \mathbb R$. The choice $\air^2 = \frac{8}{11}$ led to the near-horizon expansion of a new supersymmetric black hole, as later confirmed and studied in greater detail in \cite{Blazquez-Salcedo:2017kig, Blazquez-Salcedo:2017ghg}.
Similarly, here we can set either $\air_4=0$, or the parenthesis in \eqref{second_option_IR} to zero. Setting $\air_4=0$ leads to either (again) the solution of \cite{Gutowski:2004yv}, or to a new solution. We have integrated numerically this new solution and found that it develops a singularity in the bulk for all initial conditions we tried. So we could not find a counterpart of the regular soliton of \cite{Cassani:2014zwa} in the presence of running scalars. Thus we choose the second option to solve \eqref{second_option_IR}, that is we fix $\hir$ in terms of $\air$ as:
\begin{equation}
\label{eta_BH}
\hir = \frac{1}{48} \bigg(-8 + 11 \air^2 \pm 9 \air\sqrt{8  - 11 \air^2}\, \bigg)\ ,
\end{equation}
implying that we must take $0< \air \leq \sqrt{\frac{8}{11}}$.
Note that there are two possible values of $\eta$ depending on the sign we choose in \eqref{eta_BH}; for now we can continue by keeping this choice unspecified. Proceeding with the perturbative approach to solving the supersymmetry equation \eqref{new_eqfora} near $\rho=0$, we find that the coefficients $\air_3$ and $\air_4$ in the expansion of $a(\rho)$ remain free together with $\air$, while all the others are determined in terms of these ones. The first terms in the expansion of $a$ and $H$ read:
\begin{align}\label{aH_nearhorizon}
a \,&=\,  \air \, \rho  + \air_3\, \rho^3 + \air_4\, \rho^4 + \frac{3\air_3}{10\air}\rho^5+  \frac{\air_3 \air_4}{4\air} \rho^6+ \mathcal{O}(\rho^7)\ , \nn \\[1mm]
 H \,&=\, \hir\,\air^2 \rho^2  + 2\hir\, \air\, \air_3 \rho^4 + \frac{2\air\air_4(-2+15\air^2-24\hir)\hir}{-2+23\air^2-24\hir}\rho^5 + \frac{8\air_3^2\hir}{5}\rho^6 \, + \CO(\rho^7)\ .
\end{align}

Of the three free parameters $\air$, $\air_3$ and $\air_4$, only two are physical. Indeed it is possible to rescale at will one of the parameters, say $\air_3$, without changing the five-dimensional solution.
The reason is that eqs.~\eqref{Maxwell_orthog}, \eqref{eqfora_runningX} imply that under a rescaling of the coordinates $\rho = \lambda^{-1}\tilde \rho$, $\,y =  \lambda^2  \,\tilde y\,$, 
 a solution $a(\rho)$, $H_I(\rho)$ is transformed into another solution $\tilde a(\tilde \rho) =  \lambda\, a(\lambda^{-1}\tilde\rho)$, $\tilde H_I(\tilde \rho)= \lambda^2 H_I(\lambda^{-1}\tilde\rho)$.  
This leaves the parameters $\air$ and $\hir$ invariant, while it rescales $\air_3$, $\air_4$.
In the large-$\rho$ solution of Section~\ref{sec:near_boundary_sol}, this freedom has been fixed by assuming that for $\rho \to \infty$ the function $a$ goes like $e^{\rho}$. 
While for now we keep $\air_3$ arbitrary, when later on we will construct an interpolation between the small-$\rho$ and the large-$\rho$ solution we will need to tune it so that the assumed large-$\rho$ asymptotics are matched. So we regard $\air_3$ as an unphysical parameter.

It is also convenient to trade $\air_4$ for a new parameter $\xi$, which is invariant under such symmetry transformation and is thus physical:
\be
\air_4 = \xi \,\air_3^{3/2}.
\ee
In the following we will always use $\xi$ at the place of $\air_4$.

Notice that $\air = \sqrt{\frac{8}{11}}$ corresponds to $\eta=0$, that is $H=0$. In this case the scalar fields are fixed to their AdS value $\bar X^I$ and the gauge fields take the form \eqref{A_minimal}. This leads to a solution that is contained in minimal gauged supergravity. One can check that doing so one recovers the near-horizon expansion of the supersymmetric black hole studied in \cite{Blazquez-Salcedo:2017kig, Blazquez-Salcedo:2017ghg}. 
So we can expect that choosing $\hir$ as in \eqref{eta_BH}, but with $\air \neq \sqrt{\frac{8}{11}}$, will lead to a generalization of such black hole, where the scalars will be running. In the remainder of this section and the next ones we will show that this is indeed the case.

In the remainder of this section we provide the first terms in the small-$\rho$ expansion of the metric, the gauge fields and the scalar fields. Although these depend on the free parameters $\air$, $\air_3$, $\xi$ only, for convenience in the expressions below we also employ $\eta$, being understood that this is fixed in terms of $\air$ as in  \eqref{eta_BH}. Our main purpose will be to show that our small-$\rho$ solution has a regular horizon at $\rho = 0$.

For the function $f$ and $w$ we obtain from \eqref{f_from_a_ansatz}, \eqref{w_from_a_ansatz}:
\begin{equation}
\label{solfBH}
f = \frac{12 \, \air^2}{\Delta} \rho^2 + \frac{24 \air \air_3 \left(4 \air^2 + 
	12  \hir -1\right) \left[128 \air^4 - \left(1 - 12  \hir \right) \left(1 + 24  \hir \right) - 
	4  \air^2 \left(7 + 96 \hir \right) \right] }{\Delta^4} \rho^4 
\end{equation}
\begin{equation}
w = - \, \frac{\left(1-4 \, \air^2\right)^2-144 \, \hir ^2}{48 \, \air^2} \, \frac{1}{\rho^2} + \frac{\air_3  \left(-272  \, \air^4 + 64 \, \air^2 - 144 \, \hir^2 + 1 \right)}{24 \, \air^3} + \CO(\rho)\ ,
\end{equation}
where we have defined the quantity:
\begin{equation}
\label{Delta_BH}
\Delta = \left(4 \air^2-24 \eta -1 \right)^{1/3} \left(4 \air^2+12 \eta -1\right)^{2/3}\ .
\end{equation}

The five-dimensional metric keeping only the leading order terms in a small $\rho$ expansion then reads:
\begin{equation}
\label{metricBH_IR}
\diff s^2 = - \frac{48\air^6}{\Delta^2\Theta}\rho^4\,\diff t^2 + \Delta \bigg[  \frac{ \diff\rho^2}{12 \, \air^2 \rho^2} + \frac{1}{12} (\sigma^2_1 + \sigma^2_2) + \Theta \, \Big(\sigma_3 - \frac{2}{v^2} \diff t\Big)^2\bigg]\ ,
\end{equation}
where 
\begin{align}
\label{Coefficients_Metric_IR}
\Theta &=  \frac{16 \, \air^4 + \air^2 (8 - 96 \, \hir )- 3 (12 \, \hir + 1)^2}{48 \left(4 \, \air^2 - 24 \, \hir -1 \right)}\ .
\end{align}

It remains to determine the scalar fields and the gauge fields. 
Starting from the scalars $X^I$, we can use their expression \eqref{scalars_up_from_ansatz} to obtain:
\begin{align}
 \label{Scalars_IR_BH_up}
 X^I \! & =  \left[ \frac{\left(4 \air^2-1\right)^2-144 \hir ^2}{\Delta^2} - \frac{20736 \, \air \, \air_3 \, \hir^2 \left(4 \air^2+12 \eta -1\right)^2}{\Delta^5} \, \rho^2 \, \right] \bar{X}^I & \notag \\[1mm]
& \,  + \left[ \frac{216  \hir   \left(4 \air^2 + 12 \hir -1 \right)}{\Delta^2} - \frac{15552  \air  \air_3  \hir \left(4 \air^2 - 1 \right)\left(4 \air^2 + 12 \hir -1\right)^2}{\Delta^5}  \rho^2  \right] \bar{Y}^I\! + \CO(\rho^3). 
\end{align}
The expansion for the scalars with lower indices, $X_I$, is easily obtained from \eqref{XIdown_from_XIup}, or equivalently from \eqref{scalars_down_from_ansatz}.
The $U^I$ functions entering in the gauge fields are computed using \eqref{UI_from_ansatz} and read:
\begin{equation}
U^I = \left( \frac{4 \, \air^2 - 1}{3} + 12 \, \air \, \air_3 \, \rho ^2 \right) \bar{X}^I + 36 \, \hir \, \bar{Y}^I + \CO(\rho^3)\ .
\end{equation}
The small-$\rho$ behaviour of the gauge fields is then found to be:
\begin{align}\label{A_nearhorizon}
A^I_\psi &= \frac{\left(4 \air^2 - 36 \eta - 1 \right) \left(4 \air^2 + 12 \eta -1 \right)} {12 \left(4 \air^2 - 24 \eta -1 \right)} \,  \bar{X}^I -\frac{18 \hir \left(4 \air^2 + 12 \hir -1 \right)}{4 \air^2 - 24 \hir -1} \, \bar{Y}^I + \CO(\rho^2)\ ,\notag\\[1mm]
A^I_t & = -\frac{2}{v^2} A_\psi^I(\rho =0) + \CO(\rho^2)\ .
\end{align}

We can argue that the solution above describes the vicinity of an event horizon of finite size  situated at $\rho =0$. Indeed the elsewhere timelike supersymmetric Killing vector $V$, whose norm is $-f^2$, becomes null as $\rho \to 0$. 
Moreover the metric has a divergent term $\mathcal{O}(\rho^{-2}) \diff\rho^2$, while the remaining spatial part remains finite. 
In addition, both the scalar fields and the gauge fields have a regular behaviour as $\rho\to 0$. In particular, note that in the gauge we are using the gauge fields at the horizon are transverse to the supersymmetric Killing vector $V$,
\be\label{our_gauge}
V^\mu A^I_\mu = A^I_y = A^I_t +\tfrac{2}{v^2} A^I_\psi = 0 \quad\  \text{for}\quad \ \rho = 0\ .
\ee

\begin{figure}
	\begin{minipage}{.49\textwidth}
		\includegraphics[width=7cm]{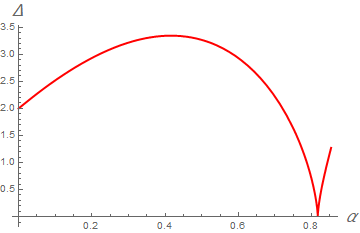}
	\end{minipage}
	\begin{minipage}{.49\textwidth}
		\includegraphics[width=7cm]{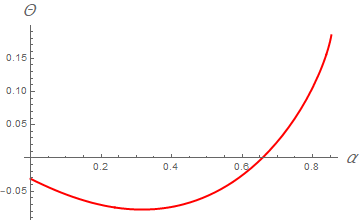}
	\end{minipage}
	\caption{The two functions $\Delta(\air)$ and $\Theta(\air)$ whose positiveness is needed to have a regular horizon. We observe that $\Delta$ is always positive except in the cusp point at $\air=\sqrt{2/3}$, while $\Theta$ is positive only for $\air \gtrsim 0.657 $.	\label{Fig:Conditions_fig}
}
\end{figure}

The geometry of the horizon is more conveniently described introducing gaussian null coordinates adapted to the supersymmetric Killing vector field $V$ \cite{Gutowski:2004ez,Gutowski:2004yv}. This is done by the transformation:
\begin{equation}
\diff y = \diff u +  \left(\frac{fw^2}{(2aa')^2}-\frac{1}{f^{2}}\right)\!\diff \tilde\rho\ , \ \quad \diff \hat\psi = \diff \tilde\psi - \frac{f\,w}{(2aa')^2} \, \diff \tilde\rho\ ,  \ \quad  \diff\rho = \, \sqrt{\frac{1}{f} - \frac{f^2w^2}{(2aa')^2}} \, \diff \tilde\rho\,,
\end{equation}
which sets the original five-dimensional metric \eqref{metric} in the form
\be
\diff s^2 = -f^2\diff u^2 +2\, \diff u\,\diff \tilde\rho - 2f^2w\,\diff u \,\tilde\sigma_3 + f^{-1}a^2 (\sigma_1^2+\sigma_2^2) + \left( f^{-1}(2aa')^2 - f^2w^2 \right)\tilde\sigma_3^2\,. 
\ee
Plugging our near-horizon solution in, we obtain that the metric at the horizon is
\be\label{HorMetric}
\diff s^2_{\rm horizon} \,=\, 2\,\diff u \,\diff \tilde\rho + \frac{\Delta}{12}(\sigma_1^2 + \sigma_2^2) + \Delta\, \Theta\, \tilde\sigma_3^2\ ,
\ee
which is manifestly well-definite and regular provided $\Delta > 0$ and $\Theta>0$. 
We have plotted these quantities in Figure \ref{Fig:Conditions_fig}, choosing the minus sign in the determination \eqref{eta_BH} for $\eta$. We note that $\Delta$ is positive for every value of the parameter $\air$ except for $\air = \sqrt{2/3}\simeq 0.816$, while $\Theta$ is real and positive for 
$ 
0.657 < \air < \sqrt{8/11} \simeq0.853.
$ 
Regularity of the horizon however does not guarantee regularity outward the horizon. In Section \ref{sec:numerics}, we will see that that regularity in the bulk in fact further constrains the allowed range of $\air$.\footnote{Similarly, we find a narrow regularity range for the horizon geometry when the plus sign is chosen in the formula \eqref{eta_BH} for $\eta$. This is also further reduced when regularity away from the horizon is imposed.}

The area of the horizon is easily computed from \eqref{HorMetric} and reads:
\be\label{AreaHorizon}
{\rm Area} = \frac{\pi^2}{3\sqrt 3}\left(4\alpha^2+12\eta-1\right)\left[16\alpha^4 + \alpha^2(8-96\eta)-3(12\eta+1)^2\right]^{1/2}\ .
\ee
This is finite in the allowed range of the parameters.

We have thus shown that our small-$\rho$ solution describes the vicinity of the horizon of a new two-parameter family of black holes with running scalars, controlled by the parameters $\air$ and $\xi$ (recall that in general our $q_I$ are not free parameters as they are fixed by condition \eqref{cond_on_q}).

We note two important facts regarding the parameter $\xi$. The first is that $\xi$ is sufficiently subleading in the small-$\rho$ expansion of $a$  not to appear in the leading terms of the supergravity fields as $\rho\to0$. In other words, the horizon is not affected by $\xi$. We will confirm later that this parameter is anyway physical, as when it is non-zero it leads to a squashing of the conformal boundary, making the solution asymptotically {\it locally} AdS (as opposed to asymptotically AdS).
The second fact is that in the limiting case $\xi = 0$ we can resum the perturbative series and obtain the exact solution $H = \eta\, a^2$, $a = \alpha \sinh \rho$, where $\eta$ is fixed in terms of $\air$ as discussed above. This matches the solution of \cite{Gutowski:2004yv}, with our parameter $\air$ being mapped into the three parameters $\alpha_1^{\rm GR}, \alpha_2^{\rm GR}, \alpha_3^{\rm GR}$ appearing in that paper. The precise relation between the parameters is easily worked out: comparing our expression \eqref{scalars_down_from_ansatz} for the scalars with the one in  \cite[eq.$\:$(3.19)]{Gutowski:2004yv}, we find the relation between our $q_I$ and the $q_I^{\rm GR}$ of~\cite{Gutowski:2004yv}:
\be
q_I^{\rm GR} = \tfrac{1}{3}(4\alpha^2-1)\ell^2 \bar{X}_I + 8\,\eta\, q_I\ ,
\ee
where we reinstated the AdS radius $\ell$.
Using the definitions given in \cite{Gutowski:2004yv}, this implies
\begin{align}
\alpha_1^{\rm GR} &= (4\alpha^2-1)\ell^2\ ,\nn\\[1mm]
\alpha_2^{\rm GR} &=  \tfrac{1}{3}(4\alpha^2-1)^2\ell^4 -48\,\eta^2 \ ,\nn\\[1mm]
\alpha_3^{\rm GR} &= \tfrac{1}{27}(4\alpha^2-1)^3\ell^6 - 16\,\eta^2 (4\alpha^2-1)\ell^2 - 128\,\eta^3 \ .
\end{align}
We have thus established that for $\xi= 0$ our solution corresponds to a one-parameter sub-family of the black hole of \cite{Gutowski:2004yv}.
Taking $\xi\neq 0$ brings us instead on a new branch of solutions. Nevertheless, since $\xi$ does not affect the horizon geometry, the latter remains the same as in the black hole of~\cite{Gutowski:2004yv}, with the identification of the parameters above. In particular, using this dictionary the area of the horizon \eqref{AreaHorizon} matches the expression given in~\cite{Gutowski:2004yv}.

Another limiting case is the one of constant scalars, obtained by taking $\air = \sqrt{\frac{8}{11}}$. We have checked that in this case our small-$\rho$ solution reduces to the one of \cite{Cassani:2014zwa,Blazquez-Salcedo:2017kig,Blazquez-Salcedo:2017ghg}, which is controlled by the one parameter $\xi$.
In this limit the scalar fields take their constant AdS$_5$ value, $X^I=\bar{X}^I$, and the part of the gauge fields along $\bar{Y}^I$ vanishes. Moreover, the horizon geometry is completely frozen. We have thus demonstrated that by allowing for running scalars one can introduce a new parameter so that the horizon geometry gets unfrozen.

\subsection{Page and Komar integrals}\label{sec:FirstIntegrals}

In this section we discuss some conserved charges that will play an important role in the following.
This generalizes to Fayet-Iliopoulos gauged supergravity similar considerations made in~\cite{Cassani:2014zwa,Blazquez-Salcedo:2017ghg} for minimal gauged supergravity.

Let us consider a Cauchy surface (that is, a hypersurface of constant time). This is foliated by three-dimensional spacelike, compact hypersurfaces of constant $\rho$, that we denote by $\Sigma_\rho$. By considering the hypersurface $\Sigma_\infty$ at $\rho=\infty$, we introduce the Page electric charges \cite{Page:1984qv}:
\be\label{Page_charges}
P_I = \frac{1}{\kappa^2} \int_{\Sigma_\infty} \left( Q_{IJ}\star F^J + \frac{1}{4}C_{IJK} A^J \wedge F^K \right)  \ .
\ee
Since by the Maxwell equation \eqref{Maxwell_general} the integrand is a closed three-form, it follows from the Stokes theorem that $P_I$  is a constant of the flow along the radial direction and can equally well be evaluated on any other hypersurface  $\Sigma_\rho$ (moreover it should be quantized in appropriate units). In particular, it can be measured at the horizon, that is on $\Sigma_{\rho =0}$.

Similarly, we can associate a conserved angular momentum to the symmetry generated by the vector $K =\frac{\partial}{\partial \psi}$ by considering the following generalization of the Komar integral:
\be\label{Komar_int}
J = \frac{1}{2\kappa^2}\int_{\Sigma_\infty} \left[\star\, \diff K  + 2\, \iota_K  A^I \left( Q_{IJ}\star F^J + \frac{1}{6}C_{IJL} A^J \wedge F^L \right)\right]\ .
\ee
 Using both the Einstein and the Maxwell equation, one can show that the integrand is closed on the Cauchy surface and thus $J$ can also be evaluated on any $\Sigma_\rho$. We emphasize that in general the standard Komar integral $\int_{\Sigma_\infty} \star \diff K$ would not satisfy this property, because of the gauge field energy-momentum tensor in the Einstein equation.

The integrals above can be expressed in a more explicit way, adapted to our supersymmetric problem. 
$P_I$ decomposes in a term proportional to $\bar X_I$ and a term proportional to $q_I$, so we can write 
\be\label{splitting_Page}
P_I= -\frac{48\pi^2\ell^2}{\kappa^2} \left(\mathcal{K}_1 \bar{X}_I + \mathcal{K}_2\, q_I\right)\ ,
\ee
where $\mathcal{K}_1$, $\mathcal{K}_2$ are two constants and the overall factor is introduced for later convenience. We also find it convenient to redefine
\be\label{defK3}
J = \frac{4\pi^2\ell^3}{\kappa^2}\mathcal{K}_3\ .
\ee
In this formulae we have reinstated the AdS radius $\ell$ to emphasize that the constants $\mathcal{K}_1,\mathcal{K}_2,\mathcal{K}_3$ are dimensionless. 
Using the supersymmetric form of the supergravity fields described in Section \ref{sec:susyeqs_general} as well as our ansatz \eqref{ansatz_H}, we find that these can be written as:\footnote{The integral over the angular coordinates yields $\int \sigma_1\wedge \sigma_2\wedge \sigma_3 = \int \sin\theta\,\diff\theta\wedge\diff\phi\wedge\diff\psi = 16\pi^2$ as we have assumed a canonical range for the Euler angles on $S^3$, $ \theta\in [0,\pi]$, $\phi\in[0, 2\pi]$ and $\psi\in [0, 4\pi]$.}
\begin{align}
\label{Maxwell_long_first_integral}
\mathcal{K}_1 \,&=\,a^3 a^\prime \left( f_{\rm min}^{-1} \right)' + \frac{1}{\ell} a^2 w  + \frac{\ell^2p^2}{18} - \frac{2}{\ell^2a^4} \, H^2  \ ,\\[1mm]
\label{Maxwell_transv_first_integral}
\mathcal{K}_2 \,&=\, H'' - \left(\frac{3a'}{a}+\frac{a''}{a'}\right) H' + \frac{2p}{3a^2}H  -\frac{4}{\ell^2a^4} H^2 \ ,\\[1mm]
\label{Einstein_First_Integral}
\mathcal{K}_3 &=\,\,\frac{a}{a^\prime f^3}  \left(f^3w^2 -4 a^2 (a^\prime)^2\right)^2 \left(\frac{f^3 w}{f^3 w^2 - 4 a^2 (a^\prime)^2}\right)' - 12 \, A^I_\psi \left(\mathcal{K}_1  \bar{X}_I + \mathcal{K}_2\, q_I \right)\nn\\[1mm]
&\ \quad + \frac{1}{3} C_{IJK} A^I_\psi A^J_\psi A^K_\psi \ .
\end{align}
Constancy of $\mathcal{K}_1$ and $\mathcal{K}_2$ immediately follows from eqs.~\eqref{MaxwellEqParallel} and \eqref{Maxwell_orthog}, which express the Maxwell equation.  
 In order to see that $\mathcal{K}_3$ is also constant one has to use the $^t{}_\psi$ component of the Einstein equation \eqref{Einstein_equations} as well as the Maxwell equation and the supersymmetry conditions. 

The quantities defined above represent a possible definition of the electric charges and the angular momentum of the solution. In Section~\ref{sec:properties_sol} we will compare them with similar quantities defined through holographic renormalization and we will also see that they are relevant for expressing the entropy of the solution.
In addition they are useful for the following more practical purpose.
In our two-parameter black hole solution, the parameters controlling the general near-boundary solution of Section~\ref{sec:near_boundary_sol} should be related to the two free parameters appearing in the near-horizon solution of Section~\ref{sec:near_horizon_sol}. Evaluating the first integrals both near the boundary and near the horizon allows to fix three of the near-boundary parameters in terms of the remaining near-boundary parameters and of the near-horizon ones.
Concretely, we evaluate~\eqref{Maxwell_long_first_integral}--\eqref{Einstein_First_Integral} at large $\rho$ using the results of Section~\ref{sec:near_boundary_sol}. We obtain three equations that can be solved for the parameters $a_4$, $H_4$ and $a_6$ appearing in the large $\rho$ solution for $a$ and $H$ as:
\begin{align}\label{a4_condition}
a_4 &= \tfrac{5}{384} + \tfrac{1}{6}a_2 -\tfrac{2}{3}a_2^2+ (1-5a_2)\tfrac{c}{12} - \tfrac{13}{48}c^2 + \tfrac{3}{8}H_2^2 + \tfrac{9}{8}H_2\HL + \tfrac{51}{64}\HL^2 - \tfrac{3}{8}\mathcal{K}_1\ ,\\[3mm]
\label{H4_condition}
H_4 &= \tfrac{1}{6} (4 a_2 H_2 + H_2 -2 \HL a_2  -4 \HL c + \HL) + H_2^2 + 2 H_2 \HL + \tfrac{3}{2}\HL^2 + \tfrac{1}{4}\mathcal{K}_2\ ,\\[3mm]
\label{a6_condition}
a_6 &= \tfrac{1}{1296} - \tfrac{5}{18}a_2^2 + \tfrac{70}{81}a_2^3 + \left(\tfrac{1913}{3888}a_2-\tfrac{125}{1944} \right)c^2 + \tfrac{1105}{11664}c^3 + \tfrac{1}{16}H_2^2 + \tfrac{1}{6}H_2^3 \nn \\[2mm]
&\quad + c\left( \tfrac{25}{3456}+\tfrac{197a_2-61}{324}a_2 - \tfrac{13}{72}H_2^2 -\tfrac{137}{216}H_2\HL - \tfrac{971}{1728}\HL^2 + \tfrac{19}{216}\mathcal{K}_1 \right)  + \tfrac{1229}{1728}\HL^3\nn \\[2mm]
&\quad + \left( \tfrac{169}{144}H_2 +\tfrac{557}{3456}\right)\HL^2  + a_2\left(-\tfrac{29}{3456}-\tfrac{17}{24}H_2^2 - \tfrac{137}{72}H_2\HL - \tfrac{2129}{1728}\HL^2 + \tfrac{43}{72}\mathcal{K}_1 \right)\nn\\[2mm]
&\quad + \HL \left( \tfrac{7}{36}H_2 + \tfrac{17}{24}H_2^2 + \tfrac{29}{288}\mathcal{K}_2 \right)-\tfrac{5}{288}\mathcal{K}_1 + \tfrac{1}{12}H_2\mathcal{K}_2 -\tfrac{1}{384}\mathcal{K}_3\ .
\end{align}
These relations hold for every asymptotic solution of the form presented in Section~\ref{sec:near_boundary_sol} and allow to eliminate $a_4$, $a_6$, $H_4$ in favour of the remaining parameters $a_0$, $a_2$, $v^2 = 1-4c$, $\HL$, $H_2$. Of course these relations also involve the integration constants $\mathcal{K}_1$, $\mathcal{K}_2$, $\mathcal{K}_3$, so we still have the same number of arbitrary parameters. However it is convenient to eliminate $a_4$, $a_6$, $H_4$ as this simplifies many expressions. Moreover this is desirable conceptually because in specific solutions the free parameters entering in the ``expectation value terms'' should be fixed in terms of the ``source terms'' by regularity conditions arising in the interior of the solution and $\mathcal{K}_1$, $\mathcal{K}_2$, $\mathcal{K}_3$ --- being independent of the radial coordinate --- are easily determined by considering the solution in the interior.  
For our black hole, they are determined by the small-$\rho$ solution given in Section \ref{sec:near_horizon_sol}, describing the vicinity of the horizon. We find that in the limit $\rho\to 0$, \eqref{Maxwell_long_first_integral}--\eqref{Einstein_First_Integral} evaluate to:
\begin{align}\label{K1_IR}
\mathcal{K}_1 &= -\frac{1}{9} \left(\air^2+1\right) \air^2+\hir ^2+\frac{5}{144}\ ,\\[1mm]
\label{K2_IR}
\mathcal{K}_2 &= -\frac{2}{3} \hir  \left(2 \air^2 + 6 \hir +1 \right)\ ,\\[1mm]
\label{K3_IR}
\mathcal{K}_3 &= -4 \left(8 \air^2+1\right) \hir ^2 + \frac{1}{108} \left(8 \air^2+7\right) \left(1-4 \air^2\right)^2  - 64  \hir ^3\ .
\end{align}
Recalling that $\hir$ is fixed as in \eqref{eta_BH}, these are functions of the near-horizon parameter $\air$ only.
In this way we have determined $a_4$, $a_6$, $H_4$ in terms of the other near-boundary parameters $a_0$, $a_2$, $v$, $\HL$, $H_2$ {\it and} the near-horizon parameter $\air$. On the other hand, in order to determine the relation of the remaining near-boundary parameters with the only two physical near-horizon parameters $\air$ and $\xi$ we will have to resort to numerics.

As a cross-check, we can evaluate the relations above in the limit leading to minimal gauged supergravity and compare with the expressions previously found within this theory \cite{Cassani:2014zwa,Blazquez-Salcedo:2017ghg}. We thus take $H_2 = H_4 = \HL = 0$. Then \eqref{H4_condition} merely gives $\mathcal{K}_2=0$, while \eqref{a4_condition}, \eqref{a6_condition} 
 reduce to expressions that are in agreement with eqs.~(4.21), (4.22) of~\cite{Blazquez-Salcedo:2017ghg}.\footnote{Upon identifying the constants $c_t,c_W$ appearing there as $c_t = -4 \sqrt{3} \mathcal{K}_1,\,  c_W = -\mathcal{K}_3$.}
The values of $\mathcal{K}_1$, $\mathcal{K}_3$ specific to the black hole solution of minimal gauged supergravity studied in~\cite{Blazquez-Salcedo:2017ghg} are correctly retrieved by sending $\air \to \sqrt{\frac{8}{11}}$ in \eqref{K1_IR}, \eqref{K3_IR}.
We can also compare with the expressions for $a_4$ and $a_6$ given in~eq.~(B.1) of \cite{Cassani:2014zwa}: we find agreement upon setting $\mathcal{K}_1=\mathcal{K}_2=\mathcal{K}_3=0$, which are the appropriate values for a solution capping off smoothly such as the one presented in that paper.

\subsection{Numerical analysis}\label{sec:numerics}

In this section we perform a numerical study showing that there is a smooth solution interpolating between the near-horizon and near-boundary regimes presented above. This happens only in a certain region of the parameter space, that we determine.

\begin{figure}[!h]
	\label{Fig:a_H_solution}
	\begin{minipage}{.49\textwidth}
		\includegraphics[width=7 cm]{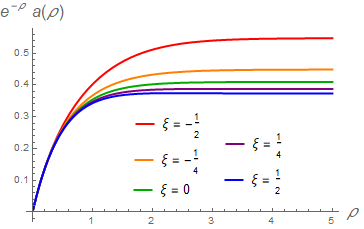}
		\text{(a) The solution $a$.}
	\end{minipage}
	\quad 
	\begin{minipage}{.49\textwidth}
		\includegraphics[width=7 cm]{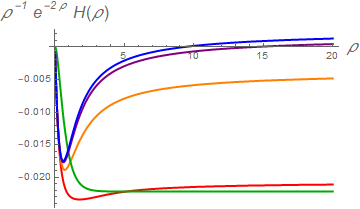}
		\text{(b) The solution $H$.}
	\end{minipage}
	\label{Fig:Metric_Components}
	\bigskip
	
	\begin{minipage}{.49\textwidth}
		\includegraphics[width=7cm]{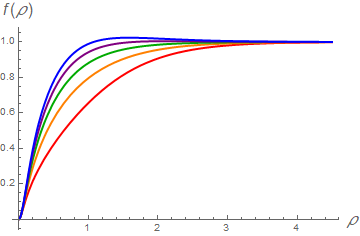}
		\text{(c) The function $f=g_{\rho \rho}^{-1}$.}
	\end{minipage}
	\bigskip
	\begin{minipage}{.49\textwidth}
		\includegraphics[width=7cm]{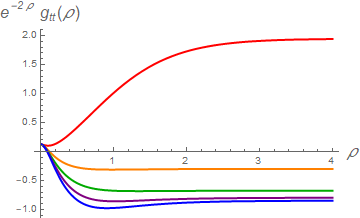}
		\text{(d) The component $g_{tt}$.}
	\end{minipage}
	\begin{minipage}{.49\textwidth}
		\includegraphics[width=7cm]{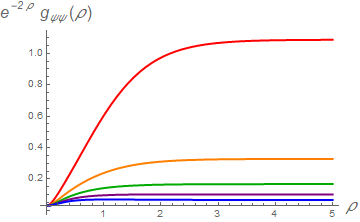}
		\text{(e) The component $g_{\psi \psi}$.}
	\end{minipage}
	\begin{minipage}{.49\textwidth}
		\includegraphics[width=7cm]{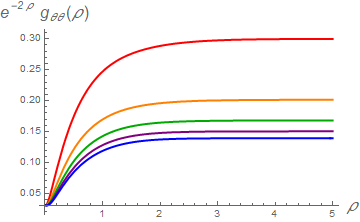}
		\text{(f) The component $g_{\theta \theta}$.}
	\end{minipage}
	\caption{	Relevant functions and metric components of our solution, rescaled by their asymptotic behaviour at large $\rho$. 
		The different values of the near-horizon parameter $\xi$ are indicated in the label.	
		We emphasize that although this is not immediately recognized from the plots, $g_{\theta \theta}$ and $g_{\psi \psi}$ go to a small but positive constant, leading to an even horizon of finite size. This is clear from~\eqref{metricBH_IR}. \label{fig:metric_plot}} 
\end{figure}

\begin{figure}[!htb]
	\label{Fig:Gauge_Field_Components}
	\begin{minipage}{.49\textwidth}
		\includegraphics[width=7cm]{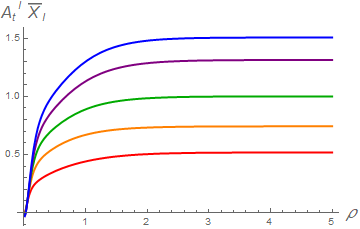}
		\begin{center}
			\text{(a) The component of $A^I_t$ along $\bar{X}^I$.}	
		\end{center}
	\end{minipage}
	\bigskip
	\begin{minipage}{.49\textwidth}
		\includegraphics[width=7cm]{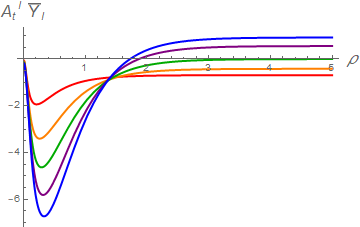}
		\begin{center}
			\text{(b) The component of $A^I_t$ along $\bar{Y}^I$.}	
		\end{center}
	\end{minipage}
	
	\begin{minipage}{.49\textwidth}
		\includegraphics[width=7cm]{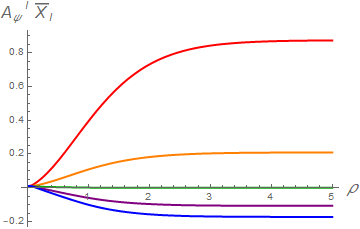}
		\begin{center}
			\text{(c) The component of $A^I_\psi$ along $\bar{X}^I$.}	
		\end{center}
	\end{minipage}
	\begin{minipage}{.49\textwidth}
		\includegraphics[width=7cm]{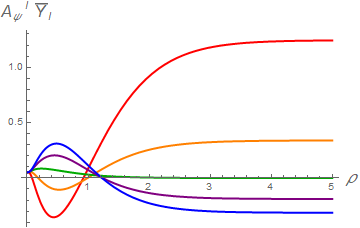}
		\begin{center}
			\text{(d) The component $A^I_\psi$ along $\bar{Y}^I$.}	
		\end{center}
	\end{minipage}
	\bigskip
	\label{Fig:Scalar_Fields_Components}
	\begin{minipage}{.49\textwidth}
		\includegraphics[width=7cm]{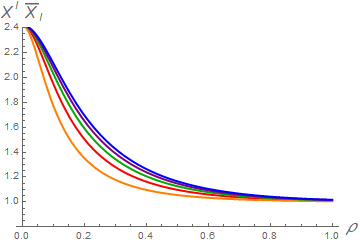}
		\begin{center}
			\text{(e) Scalar fields $X^I$ along $\bar{X}^I$. }	
		\end{center}
	\end{minipage}
	\bigskip
	\begin{minipage}{.49\textwidth}
		\includegraphics[width=7cm]{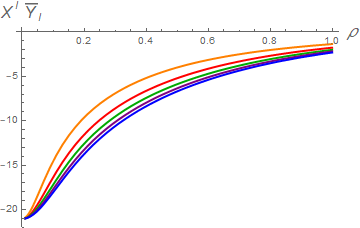}
		\begin{center}
			\text{\quad (f) Scalar fields $X^I$ along $\bar{Y}^I$.}	
		\end{center}
	\end{minipage}
	\caption{Components of the gauge fields $A^I$ and of the scalar fields $X^I$ along $\bar{X}^I$ and~$\bar{Y}^I$. 
		\label{fig:scalar_plot}
	}
\end{figure} 
\begin{figure}
 	\label{Fig:Squash_Xi}
 	\centering
 	\includegraphics[width= 8 cm]{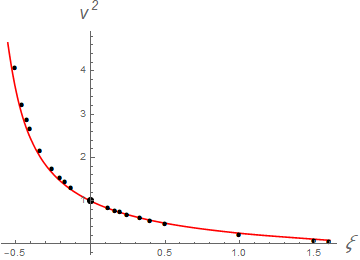}
 	\caption{Relation between the near-horizon parameter $\xi$ and the squashing $v^2$ of the boundary metric, for $\air = 0.82$.  $v^2$ is positive and finite for $-0.7 \lesssim \xi \lesssim 1.6$. The black dots represent the values effectively calculated by means of the numerical analysis. The larger dot at $(\xi = 0,v=1)$ represents the solution of \cite{Gutowski:2004yv}.\label{fig:xi_v_plot}}
\end{figure}
\begin{figure}
	\label{Fig:UV_a_param_map}
	\begin{minipage}{.50\textwidth}
		\includegraphics[width=7.2cm]{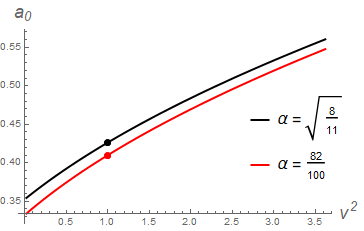}
		\begin{center}
			\text{(a) The parameter $a_0$.}	
		\end{center}
	\end{minipage}
	\bigskip
	\begin{minipage}{.48\textwidth}
		\includegraphics[width=6.4cm]{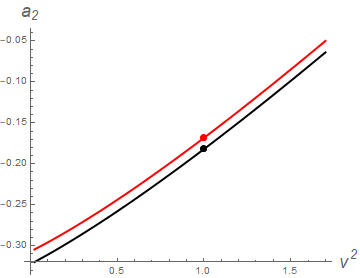}
		\begin{center}
			\text{(b) The parameter $a_2$.}
		\end{center}
	\end{minipage}
	\begin{minipage}{.49\textwidth}
		\includegraphics[width=7cm]{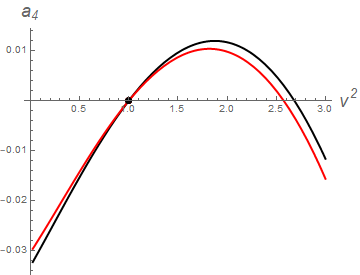}
		\begin{center}
			\text{(c) The parameter $a_4$.}
		\end{center}
	\end{minipage}
	\begin{minipage}{.49\textwidth}
		\includegraphics[width=7cm]{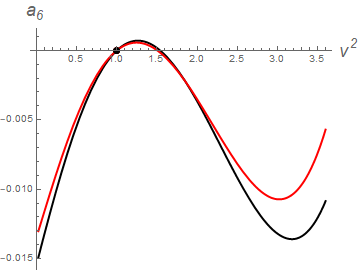}
		\begin{center}
			\text{(d) The parameter $a_6$.}
		\end{center}
	\end{minipage}
	\caption{The near-boundary parameters $a_0,a_2,a_4,a_6$ in terms of the squashing $v^2$, for $\air = 0.82$ (red) and $\air = \sqrt{8/11}$ (black). \label{fig:a_v_plot}}
\end{figure}
\begin{figure}
	\begin{minipage}{.49\textwidth}
		\includegraphics[width=7cm]{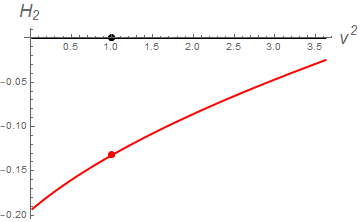}
		\begin{center}
			\text{(a) The parameter $H_2$.}	
		\end{center}
	\end{minipage}
	\begin{minipage}{.49\textwidth}
		\includegraphics[width=7cm]{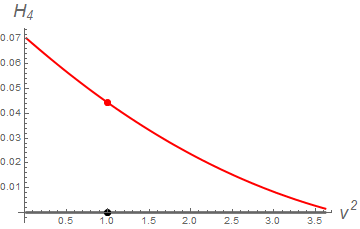}
		\begin{center}
			\text{(b) The parameter $H_4$.}
		\end{center}
	\end{minipage}
	\begin{minipage}{.49\textwidth}
		\includegraphics[width=7cm]{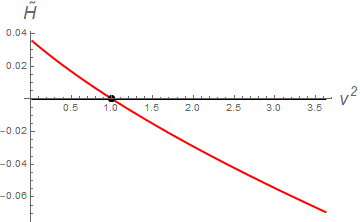}
		\begin{center}
			\text{(c) The parameter $\HL$.}
		\end{center}
	\end{minipage}
	\caption{The parameters of $H_2, H_4, \HL$ in terms of the squashing $v^2$, for $\air = 0.82$ (red). For $\air = \sqrt{\frac{8}{11}}$ they vanish identically (black).\label{fig:H_v_plot}}
\end{figure}

We start by briefly describing how we perform the numerical analysis. We fix the initial conditions at $\rho \simeq0$ using the expressions in Section~\ref{sec:near_horizon_sol} and integrate equations \eqref{eqforH}, \eqref{new_eqfora} numerically towards larger values of~$\rho$.
Of course, in order to do this we need to assign a numerical value to the two physical parameters $\xi$ and $\air$. In Section~\ref{sec:near_horizon_sol} we saw that regularity of the horizon (for the minus sign choice in \eqref{eta_BH}) requires $0.657 \leq \air \leq \sqrt{\frac{8}{11}}$ and $\air\neq\sqrt{\frac{2}{3}}$, so we perform our analysis for various values of $\air$ within this range.
 Moreover we rescale the unphysical parameter $\air_3$ in such a way that the assumed AlAdS behaviour of $a$ for $\rho \to \infty$ holds.\footnote{In order to achieve this we exploit the rescaling properties described under eq.~\eqref{aH_nearhorizon}. We integrate a first time choosing $\air_3 = 1$, then we look at the large-$\rho$ behaviour of the solution and determine the rescaling factor $\lambda^2$ by requiring that $f\to 1$  asymptotically. This is equivalent to impose $a \sim e^\rho$ as $\rho\to\infty$. Then we fix $\air_3 = 1/\lambda^2$ and repeat the integration. 
}

The numerical analysis shows that the solution is regular only in the range:
\begin{equation}
\label{air_1_range}
\sqrt{\frac{2}{3}} < \air \leq \sqrt{\frac{8}{11}}\ ,
\end{equation} 
while outside of this the function $f$ presents a divergence at finite $\rho$ and the same do other components of the metric and the gauge fields. We have checked for several values of $\air$ within this range that all the components of the metric and the gauge fields are regular, provided $\xi$ lies in a certain range that depends on $\air$ and is determined by regularity of the boundary geometry. 

We report as an illustrative example the relevant physical functions for the value $\air = 0.82$ and for different choices of $\xi$. In Figure~\ref{fig:metric_plot} we display the functions $a$ and $H$ and the components of the metric \eqref{comp_5dmetric}, while
 Figure \ref{fig:scalar_plot} shows the components of the gauge field \eqref{comp_5dgaugefield} and of the scalar fields $X^I$. The plots demonstrate that the solution is smooth on and outside the event horizon. 

Our next goal is to determine the free parameters appearing in the general near-boundary solution ($a_0$, $a_2$, $a_4$, $a_6$, $v$, $H_2$, $H_4$, $\HL$) as functions of the only two near-horizon parameters  $\air$, $\xi$ characterizing the black hole solution. In order to do this we compare the numerical solution for the functions $a$ and $H$ with the  near-boundary expansion discussed in Section~\ref{sec:near_boundary_sol} at some reasonably large values of the radial coordinate $\rho$ (we find it sufficient to use several points in the interval $3 < \rho <6$), and evaluate the near-boundary parameters using a best-fit technique. In Figures~\ref{fig:xi_v_plot},~\ref{fig:a_v_plot},~\ref{fig:H_v_plot} we present the results obtained 
using this method for the two values $\air = 0.82$ and $ \alpha_1= \sqrt{\frac{8}{11}}$ and for about $20$ values of $\xi$. Figure~\ref{fig:xi_v_plot} shows the relation between the squashing parameter $v^2$ and the near-horizon parameter $\xi$, with $\air = 0.82$ (we are not presenting the plot for $\air=\sqrt{\frac{8}{11}}$ as it is not significantly different from the displayed one). Notice that for $\xi$ running between $\xi \sim 1.6$ and $\xi \sim -0.7$ the squashing $v^2$ spans the whole positive line. From an AdS/CFT perspective, the squashing parameter $v$ of the boundary geometry seems to play a more significant role than $\xi$, so once $\air$ has been fixed, we choose to regard the family of solutions as parametrized by $v^2$ rather than $\xi$. Consequently, in the Figures~\ref{fig:a_v_plot} and \ref{fig:H_v_plot} we plot the near-boundary parameters as function of $v^2$. 
Recall that the solution with $\air= \sqrt{\frac{8}{11}}$ fits into minimal gauged supergravity and coincides with the black hole of \cite{Blazquez-Salcedo:2017ghg}, so with the plots of Figures~\ref{fig:a_v_plot} and \ref{fig:H_v_plot} we are comparing our new family of solutions with that one.\footnote{For the solution in minimal gauged supergravity, the plot of $a_6$ corrects the one in Figure 14 of \cite{Blazquez-Salcedo:2017ghg}. We thank the authors of \cite{Blazquez-Salcedo:2017ghg} for correspondence on this.}

With the help of the figures we can discuss some physical properties of our solution. From Figure~\ref{fig:metric_plot} we can exclude the presence of closed timelike curves, which would appear whenever the $g_{\psi \psi}$ component of the metric becomes negative. 
Although the figure displays just the behavior for $\air = 0.82$, we have verified that closed timelike curves are also absent for different values of $\air$ in the range \eqref{air_1_range}.
Furthermore we should note from Figure~\ref{fig:metric_plot} that in the near-horizon region $g_{tt}$ becomes positive, implying that the vector $\frac{\partial}{\partial t}$ becomes spacelike. This means that if this vector is regarded as the generator of time translations, then our solution presents an ergoregion for all the values of $\xi$ and $\air$ in the allowed range \eqref{air_1_range}. However we may also take as generator of time translations the supersymmetric Killing vector field \eqref{Killing_tpsi}, which corresponds to working in a frame that is co-rotating with the event horizon.  
In this case there is no ergoregion as this vector is timelike everywhere outside the horizon. This feature is common in rotating, asymptotically AdS black holes and in the supersymmetric context it was noted in \cite{Gutowski:2004ez}.

Recall that in Section~\ref{sec:FirstIntegrals} we exploited three first integrals of the equations of motion and solved for $a_4$, $a_6$, $H_4$ in terms of the other near-boundary parameters and the near-horizon parameter $\air$.  
We have checked that the values of the parameters extracted numerically are in excellent agreement with these relations.

\begin{figure}
	\centering
	\includegraphics[width= 11.5 cm]{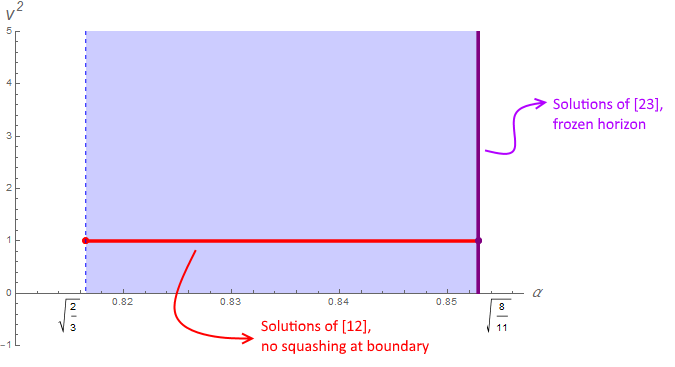}
	\caption{Parameter space of our solution. The range of the near-horizon parameter is $\sqrt{2/3}<\air\leq \sqrt{8/11}$, while for squashing at the boundary we have $0< v^2<\infty$. For $v^2=1$ we recover a sub-family of the solution of \cite{Gutowski:2004yv}, while for $\alpha=\sqrt{8/11}$ we reduce to the solution of \cite{Blazquez-Salcedo:2017ghg}. \label{Fig:ParameterSpace}}
\end{figure}

In Figure~\ref{Fig:ParameterSpace} we provide a summarizing plot of the parameter space of our solution including its notable limits.
%

%%%%%%%%%%%%%%%%%%%%%%%%%%%%%%%%%%%%%%%%%%%%%%%%%%%%%%%%%%%%%%%%%%%%%%%%%%%%%%%%%%%%%%
\section{Holographic renormalization and physical properties}\label{sec:properties_sol}

We have established that our black hole solution is controlled by two parameters. One of the two (it can be seen as $v$) does not affect the near-horizon geometry but introduces a non-trivial squashing of the $S^3$ at the conformal boundary. The solution is therefore AlAdS: it is only when the $S^3$ is round ($v=1$) that a conformally flat boundary is obtained. In this section we evaluate the conserved charges, the on-shell action as well as the entropy of the solution and discuss the relations between such quantities. While for non conformally flat boundaries the Ashtekar-Das method \cite{Ashtekar:1999jx} for computing conserved charges does not apply, we can resort to holographic renormalization. By introducing suitable boundary counterterms, holographic renormalization removes the large-distance divergences that are encountered in AlAdS spaces and in this way provides well-defined energy-momentum tensor and currents. The latter have a natural interpretation as one-point functions of the holographically dual field theory operators. Some general references on holographic renormalization that are also relevant for our problem are~\cite{Witten:1998qj,Henningson:1998gx,Balasubramanian:1999re,deHaro:2000vlm,Bianchi:2001kw,Martelli:2002sp,Papadimitriou:2005ii}.

\subsection{Holographic renormalization in Fayet-Iliopoulos gauged supergravity}\label{sec:HoloRenoFIsugra}

We start by providing some general formulae for holographic renormalization in five-dimensional Fayet-Iliopoulos gauged supergravity. These will be valid under the assumption that the fermion fields are set to zero and that the scalar fields only depend on the radial coordinate.

We find it convenient to present the results of this section using the Fefferman-Graham radial coordinate $r$ introduced in Appendix \ref{app:FeffermanGraham}. Although we could equally well work with the original coordinate $\rho$, the choice of $r$ is more standard in holography and may facilitate comparison with other references. 
We recall that the general Fefferman-Graham form of the five-dimensional metric is:
\begin{equation}
\label{FGmetric}
\diff s^2 = \ell^2\frac{\diff r^2}{r^2} + h_{ij} (x,r) \, \diff x^i \, \diff x^j\ ,
\end{equation}
where we have reinstated the AdS radius $\ell$ that was set to unity in the previous section. The five-dimensional spacetime $M$ is foliated by timelike hypersurfaces of constant $r$, parameterized by coordinates $x^i$, $i=0,\ldots,3$.
The asymptotic expansion of the induced metric $h_{ij}$ and the other supergravity fields is (see Appendix~\ref{app:FeffermanGraham} for more details):
\begin{equation}
\label{FGmetricomponents_maintext}
h_{ij}(x,r) \,=\, \frac{r^2}{\ell^2} \, h_{ij}^{(0)} + \ldots\ ,
\end{equation}
\begin{equation}
\label{FGgauge_maintext}
A_i^I(x,r) \,=\, A_i^{I \, (0)} + \frac{A_i^{I \, (2)} + \tilde{A}_i^{I \, (2)} \, \log{\tfrac{r^2}{\ell^2}}}{(r/\ell)^2} + \dots\ ,
\end{equation}
\begin{equation}
\label{FGscalars_up_maintext}
X^I = \, \bar{X}^I + \frac{\phi^{I \, (0)} + \tilde{\phi}^{I \, (0)} \, \log{{\tfrac{r^2}{\ell^2}}}}{(r/\ell)^2} + \dots \ ,
\end{equation}
where the leading terms $h^{(0)}_{ij}$, $A^{I\,(0)}_i$, $\tilde\phi^{I\,(0)}$ are the metric, gauge fields and scalar fields induced on the conformal boundary $\partial M$. These are interpreted holographically as background fields for the dual SCFT.

It is useful to illustrate how these background fields are organized in four-dimensional supersymmetry multiplets.
On general grounds, the bulk supergravity transformations that preserve the Fefferman-Graham gauge induce the transformations of four-dimensional conformal supergravity at the boundary.\footnote{See~e.g.~\cite{Freedman:2012zz} for an account of conformal supergravity in four dimensions.} The asymptotic values of the Fayet-Iliopoulos gauged supergravity fields provide the boundary Weyl multiplet, whose physical bosonic fields are the four-dimensional metric and an Abelian gauge field, and $n_V$ vector multiplets, whose bosonic fields are a vector and a D-term. From the point of view of the dual $\mathcal{N}=1$ SCFT, these are background multiplets sourcing the energy-momentum tensor multiplet and $n_V$ Abelian flavour current multiplets.
Specifically, since the gauge field entering in the bulk gravitino variation is $\bar{X}_I A^I$, its boundary value $\bar{X}_I A^{I\,(0)}$ should be identified with the gauge field belonging to the background Weyl multiplet and sourcing the dual R-current. On the other hand, $\bar{Y}_I A^{I\,(0)}$ and the boundary scalar field $\bar{Y}_I\, \tilde\phi^{I\,(0)}$ belong to a background vector multiplet and source the current and the scalar operator with conformal dimension $\Delta=2$ in the dual $\mathcal{N}=1$ flavour current multiplet. 

In particular, we can consider the supergravity model with $n_V=2$ arising as a consistent truncation of type IIB supergravity on $S^5$ (summarized at the end of Section \ref{sec:FYsugra}) and its dual $\mathcal{N}=4$ super Yang-Mills theory. In this case the  field theory operators $\mathcal{O}_I$ sourced by the $\tilde{\phi}^{I(0)}$, $I=1,2,3$, are identified as follows. Start from the adjoint scalars $z_i$, $i=1,\ldots,6$, in the $\mathcal{N}=4$ Yang-Mills multiplet and build the $\Delta=2$ operators ${\rm Tr}(z_iz_j-\frac{1}{6}\delta_{ij}z_kz_k)$, transforming in the ${\bf 20}'$ of $\SO(6)$. Then restrict to the singlets under $\U(1)^3\subset \SO(6)$. These may be taken as: $\mathcal{O}_1 = \frac{1}{3}{\rm Tr}\left(2 z_1^2 + 2 z_2^2 - z_3^2 - z_4^2 - z_5^2-z_6^2\right)$, $\mathcal{O}_2 = \frac{1}{3}{\rm Tr}\left(2 z_3^2 + 2 z_4^2 - z_5^2 - z_6^2 - z_1^2-z_2^2\right)$,  
$\mathcal{O}_3=-\mathcal{O}_1-\mathcal{O}_2$.
Our solution has a source term $\bar Y^I \mathcal{O}_I$. Since for the supergravity theory dual to $\mathcal{N}=4$ super Yang-Mills we need to fix $\bar Y^1=\bar Y^2=-\frac{1}{2}\bar Y^3$ (or cyclic permutations of this, recall the observation under \eqref{scalars_up_from_ansatz}), we conclude that precisely one of the $\mathcal{O}_I$ operators is sourced.

Having discussed what are the relevant SCFT background fields, we can now proceed to compute the one-point functions for the corresponding SCFT operators. In order to do this we need to set up holographic renormalization for Fayet-Iliopoulos supergravity.

 In the Fefferman-Graham gauge, the hypersurfaces of constant $r$ are homeomorphic to the conformal boundary, which is found at $r\to \infty$.
In order to regulate the large-distance divergences that appear when evaluating the supergravity action one imposes a cutoff $r_0$, so that the solution extends only up to $r=  r_0$. We denote by $M_{r_0}$ the regulated spacetime and by $\partial M_{r_0}$ its boundary at $r=r_0$.
Holographic renormalization consists of introducing appropriate local counterterms on $\partial M_{r_0}$ such that the large-distance divergences are cancelled once the cutoff is removed by sending $r_0\to\infty$. 
The renormalized action is defined as
\be
S_{\text{ren}} = \lim_{r_0 \to \infty} S_\text{reg}\ ,
\ee
where the regularized (and subtracted) action $S_\text{reg}$ is
\be
S_{\text{reg}} =  S_\text{bulk} + S_{\text{GH}} + S_{\text{ct}} \ .
\end{equation}
Here, $S_\text{bulk}$ denotes the bulk supergravity action \eqref{Bulk_action} evaluated on $M_{r_0}$. The second term is the Gibbons-Hawking boundary integral, which makes the Dirichlet variational problem for the metric well-defined. It reads:
\begin{equation}
\label{Gibbons_Hawking_action}
S_{\text{GH}} = \frac{1}{\kappa^2} \int_{\partial M_{r_0}} {\diff^4 x \, \sqrt{h} \, K}\ ,
\end{equation}
where $K=h^{ij}K_{ij}$ is the trace of the extrinsic curvature $K_{ij} 
 = \frac{r}{2\ell}\frac{\partial{h_{ij}}}{\partial r}$ of $\partial M_{r_0}$, and $h=|\det h_{ij}|$. 
Finally, $S_\text{ct}$ consists of the counterterms needed to cancel the divergences of $S_\text{bulk} + S_{\text{GH}}$. These are local boundary terms that should preserve the relevant symmetries and may contain finite contributions in addition to divergent terms.
 Although the full set of counterterms does not seem immediately available in the literature for solutions to Fayet-Iliopoulos gauged supergravity where both the scalar and the gauge fields are running and have their leading asymptotic modes turned on,  it is not hard to generalize the counterterms given in Section 5.1 of \cite{Bianchi:2001kw} (also using the results of \cite{Kikuchi:2015hcm,An:2017ihs}) to our setup. This leads us to:
\begin{align}
\label{Counterterm_action}
S_\text{ct} = - \frac{1}{\kappa^2} \int_{\partial M_{r_0}} \diff^4 x \, \sqrt{h} \,\bigg[& \CW \, +  \, \Xi \, R - \frac{\left(\CW - 3\ell^{-1} \right)}{ \log {\frac{r_0^2}{\ell^2}}} + \notag \\ 
& + \frac{\ell^3}{16}  \log{\frac{r_0^2}{\ell^2}} \left(R_{ij} R^{ij} - \frac{1}{3} R^2 - 2\ell^{-2} \, Q_{IJ}  F^I_{i j}  F^{J \, i j} \right) \bigg]\ .
\end{align}
 In this formula, the Ricci tensor $R_{ij}$ and the Ricci scalar $R$ are those of the induced metric $h_{ij}$, which is also used to raise the indices. The other ingredients are the field strengths $F^I_{ij}$ on $\partial M_{r_0}$ and two real functions of the scalar fields: the superpotential $\CW$ and the function $\Xi$. The superpotential can be read from the supersymmetry variation of the gravitino field and satisfies
\be\label{relPotSuperpot}
\mathcal{V} = \frac{1}{2}\left(Q^{IJ} -\frac{2}{3}X^IX^J\right) \partial_I\CW \, \partial_J\CW - \frac{2}{3}\CW^2\ ,
\ee
where $\mathcal{V}$ is the scalar potential. For our Fayet-Iliopoulos gauging with scalar potential \eqref{scalarpot}, the superpotential reads:
\be\label{Superpotential}
\CW = 3\,\ell^{-1} \bar{X}_I  X^I\ .
\ee
For the function $\Xi$ we may take instead:
\begin{equation}
\label{Xi_Function}
\Xi = \frac{\ell}{4} \,\bar{X}^I X_I \ .
\end{equation} 
Note that this is proportional to the scalar potential. 
At large $r_0$, it reads
$\Xi = \frac{\ell}{4} + \CO \left(r_0^{-4}  \right)$ while $\sqrt{h}R = \mathcal{O}(r_0^2)$, hence the only term in $\Xi$ that contributes to $S_{\rm ct}$ after removing the cutoff is the leading one.\footnote{The relation between the scalar potential $\mathcal{V}$ and the superpotential $\mathcal{W}$ is usually given in terms of the physical scalars $\Phi^a$, $a=1,\ldots,n_V$ and their inverse kinetic matrix $\mathcal{G}^{ab}$ as $$\mathcal{V} = \frac{1}{2}\mathcal{G}^{ab}\partial_a \CW \partial_b \CW - \frac{2}{3}\CW^2\,.$$
However in our parameterization of the five-dimensional supergravity scalar manifold, one can show that $\mathcal{G}^{ab}\partial_a X^I \partial_b X^J= Q^{IJ} -\frac{2}{3}X^IX^J$ \cite{Behrndt:1998jd} and in this way reach \eqref{relPotSuperpot}. The equation that determines $\Xi$ can be found in \cite{Batrachenko:2004fd} (see also \cite{Kikuchi:2015hcm,An:2017ihs} for more general analyses) and reads in our notation:
\begin{equation*}
\frac{2}{3}\,\Xi - \frac{1}{\CW} \mathcal{G}^{ab}\,\partial_a \Xi \, \partial_b \mathcal{W} - \frac{1}{2\mathcal{W}}=0\ .
\end{equation*}
It is not hard to see that \eqref{Xi_Function} does solve it.
}

The counterterms \eqref{Counterterm_action} cancel all divergences from $S_\text{bulk} + S_{\text{GH}}$. Specifically, the first two terms are local covariant expressions on $\partial M_{r_0}$ which remove power-law divergences, while the other terms explicitly depend on the cutoff and cancel logarithmic divergences. In addition, the first line of \eqref{Counterterm_action} yields finite terms that play an important role in the evaluation of the holographic correlation functions.

From the renormalized action one can obtain the holographic one-point functions of the energy-momentum tensor, the electric currents and the relevant scalar operators in the field theory states dual to the supergravity solution of interest.

The holographic energy-momentum tensor is defined as:
\begin{equation}
\label{Stress_Tensor_QFT}
\braket{T_{ij}} = -   \frac{2}{\sqrt{h^{(0)}}} \, \frac{\delta S_\text{ren}}{\delta h^{ij(0)}}  = - \lim_{r_0 \to \infty} \frac{r_0^2}{\ell^2} \, \frac{2}{\sqrt{h}} \, \frac{\delta S_\text{reg}}{\delta h^{ij}}\ .
\end{equation} 
Starting from the action defined above we obtain:
\begin{align}
\label{Stress_Tensor_QFT_explicit}
\braket{T_{ij}} = - \frac{1}{\kappa^2} \, \lim_{r_0 \to \infty} \frac{r_0^2}{\ell^2} &\, \bigg[  K_{ij} - K \, h_{ij} + \CW \, h_{ij} - \frac{\CW - 3\ell^{-1}}{\log{\frac{r_0^2}{\ell^2}}} \, h_{ij} - 2\,  \Xi \left(R_{ij} - \frac{1}{2} \, R \, h_{ij} \right) \notag \\[1mm]
& - \frac{\ell^3}{4}\log{\frac{r_0^2}{\ell^2}} \left(- \frac{1}{2} \, B_{ij} - \frac{2}{\ell^2} \, Q_{IJ}  F^I_{ik}  F^{J}{}_{\!j}{}^{k} + \frac{1}{2 \ell^2} \, h_{ij} \, Q_{IJ}  F^I_{kl}  F^{J \, kl} \right) \bigg],
\end{align} 
where the Ricci tensor $R_{ij}$, the Ricci scalar $R$ and the Bach tensor $B_{ij}$ are those of the induced metric $h_{ij}$ on $\partial M_{r_0}$, which is also used to raise the indices  (see e.g.~\cite{Cassani:2014zwa} for more details on the Bach tensor and how it arises here).  The contributions from the variation of the counterterm action cancel all divergences, including the logarithmic ones, so that $\braket{T_{ij}}$ is finite in the limit.

The holographic electric current is defined by varying the action with respect to the gauge field at the boundary:
\begin{equation}
\langle j^i_I \rangle  =   \frac{1}{\sqrt{h^{(0)}}} \frac{\delta S_{\text{ren}}}{\delta A_i^{I(0)}} =  \lim_{r_0\to\infty} \frac{r_0^4}{\ell^4} \,\frac{1}{\sqrt{h}} \frac{\delta S_{\text{reg}}}{\delta A_i^I}\ .
\end{equation}
We obtain:
\begin{align}\label{Holo_current}
\langle j^i_I \rangle &= - \frac{1}{\kappa^2}\lim_{r_0\to\infty} \frac{r_0^4}{\ell^4}\left[\tfrac{1}{6}\epsilon^{ijkl} \left( Q_{IJ}\star\! F^J + \tfrac{1}{6}C_{IJK} A^J \wedge F^K  \right)_{jkl} +\ell \,\nabla_j \big(Q_{IJ}F^{J\,ji}\big)\log \frac{r_0}{\ell} \right]\nn\\[1mm]
&= - \frac{1}{\kappa^2}\left[ 2\,\bar{Q}_{IJ} \left( A^{Ji\,(2)} + \tilde A^{Ji\,(2)} \right) + \tfrac{1}{12} C_{IKL}\, \epsilon^{ijkl \,(0)} A_j^{K\,(0)} F_{kl}^{L\,(0)} \right]\, ,
\end{align}
where in the first line the supergravity fields on $\partial M_{r_0}$ appear, while in the second line we have evaluated the limit and expressed the result using the Fefferman-Graham expansion \eqref{FGmetricomponents_maintext}, \eqref{FGgauge_maintext}. 
 From a dual $\mathcal{N}=1$ superconformal field theory perspective, $\bar X^I j_I$  corresponds to the R-current while the orthogonal projections correspond to $n_V$ Abelian flavour currents. 

The one-point function of the scalar operators is defined as:
\be
\langle \mathcal{O}_I\rangle =  \frac{1}{\sqrt{h^{(0)}}} \frac{\delta S_{\rm ren}}{\delta \tilde\phi^{I(0)}}  =\lim_{r_0\to\infty}\left(\frac{r_0^2}{\ell^2}\log \frac{r_0^2}{\ell^2}\,\frac{1}{\sqrt{h}} \frac{\delta S_{\rm reg}}{\delta X^{I}}\right)\ ,
\ee
where it is understood that the variation respects the constraint \eqref{constraint}, which implies $\bar{X}_I \,\delta \tilde\phi^{I(0)}=0$. 
 By going through the computation, we arrive at:
\be\label{vev_O_general}
\langle\mathcal{O}_I\rangle = \frac{2}{\kappa^2} \,\bar{Q}_{IJ}\, \phi^{J\,(0)} \ ,
\ee
where we recall that $\phi^{(0)}$ is the $\mathcal{O}(r^{-2})$ term in the Fefferman-Graham expansion \eqref{FGscalars_up_maintext} of the  scalar fields. As anticipated, this term describes the expectation value of the dual field theory operators, and here we have provided the precise relation between the two.

We remark that the formulae \eqref{Stress_Tensor_QFT_explicit}, \eqref{Holo_current}, \eqref{vev_O_general} hold for any AlAdS solution to five-dimensional Fayet-Iliopoulos gauged supergravity, under the assumption that the fermion fields are set to zero and the scalars are independent of the boundary coordinates (otherwise we would have additional terms). 

The one-point functions above satisfy the following Ward identities involving the boundary fields $h^{(0)}_{ij}$, $A_i^{I\,(0)}$, $\tilde\phi^{(0)I}$:
\begin{align}
&\nabla_i\langle j^i_I \rangle = \mathcal{A}^{\rm chiral}_I\ ,\label{conservation_current}\\[2mm]
&\nabla^{i}\langle T_{ij} \rangle  = F^{I(0)}_{ji} \langle j^i_I \rangle - A^{(0)}_j \nabla_i\langle j^i_I \rangle \ , \label{conservation_enmomtensor} \\[2mm]
& \braket{T_i{}^{i}} - 2\tilde\phi^{I(0)}\langle \mathcal{O}_I \rangle \,=\, \mathcal{A}^{\rm Weyl} \ \label{conformal_Ward}\ ,
\end{align}
where the indices are raised and the covariant derivatives are defined using $h^{(0)}_{ij}$.
These Ward identities are obtained by studying the variation of the renormalized action under gauge transformations, diffeomorphisms and conformal transformations at the boundary, respectively.\footnote{In particular, if $\delta\sigma$ is an infinitesimal conformal factor, conformal transformations act on the boundary fields as $\delta h^{(0)}_{ij}=2h^{(0)}_{ij}\delta \sigma$, $\delta A^{I(0)}=0$, $\delta \tilde \phi^{I(0)}= -(d-\Delta) \tilde \phi^{I(0)} \delta \sigma = -2  \tilde \phi^{I(0)} \delta \sigma$.}
The terms $\mathcal{A}^{\rm chiral}_I$ and $\mathcal{A}^{\rm Weyl}$ express the chiral and Weyl anomalies of the dual field theory. The former reads:
\be
\mathcal{A}^{\rm chiral}_I = - \frac{1}{24\kappa^2} C_{IKL}\, \epsilon^{ijkl \,(0)} F_{ij}^{K\,(0)} F_{kl}^{L\,(0)}\ ,
\ee
while
$\mathcal{A}^{\rm Weyl}$ is computed by taking the limit:
\be\label{anomaly as limit}
\mathcal{A}^{\rm Weyl} = \frac{1}{\kappa^2}\lim_{r_0\to\infty}\frac{r_0^4}{\ell^4} \left[ \frac{\ell^3}{8}\! \left(R_{ij} R^{ij} - \tfrac{1}{3} R^2 - 2\ell^{-2} \, Q_{IJ}  F^I_{i j}  F^{J \, i j} \right)   + 2\!\left(\CW - 3\ell^{-1} \right)\! \left(\log {\tfrac{r_0^2}{\ell^2}}\right)^{\!-2} \right]
\ee
which yields:
\be\label{Weyl_anomaly}
\mathcal{A}^{\rm Weyl} =  \frac{\ell^3}{8\kappa^2}\left[\left(R_{ij} R^{ij} - \tfrac{1}{3} R^2 - 2\ell^{-2} \, \bar Q_{IJ}  F^I_{i j}  F^{J \, i j}\right)^{(0)} + 16\ell^{-4}\,\bar Q_{IJ}\tilde\phi^{(0)I}\tilde\phi^{(0)J} \right]\ ,
\ee
where the suffix (0) indicates that now all quantities are evaluated at the conformal boundary $\partial M$.
 It may be useful to observe that the two terms in \eqref{anomaly as limit} contribute with an opposite relative sign compared to their appearance in the logarithmic divergence of the counterterm action \eqref{Counterterm_action}.  
Therefore such divergence is not the same as the Weyl anomaly. As explained in \cite{Martelli:2002sp}, this is a general feature in the presence of scalar fields dual to operators of conformal dimension $\Delta = d/2$ (where $d$ is the dimension of the boundary), as it is the case for us.

Before moving on to evaluate the formulae above in our setup let us comment on the renormalization scheme adopted. The counterterms in \eqref{Counterterm_action} cancelling power-law divergences are gauge invariant and covariant on $\partial M_{r_0}$. A priori of other symmetries, one could define a different renormalization scheme by adding finite counterterms constructed using the boundary fields. In the present context however we are interested in a supersymmetry-preserving scheme, so most of such terms would not be allowed. The issue of a supersymmetry-preserving renormalization scheme is particularly subtle in AdS$_5$/CFT$_4$. It was pointed out in \cite{Cassani:2014zwa} and further shown in \cite{Genolini:2016sxe,Genolini:2016ecx} that the scheme above does not respect the dual field theory supersymmetric Ward identities in curved space, already in the case where no supergravity vector multiplets are introduced. In \cite{Papadimitriou:2017kzw,An:2017ihs} this violation was understood as an anomaly arising in the supersymmetry transformation of the SCFT supercurrent. The anomaly affects the superalgebra in curved space and thus the BPS relation between the charges of supersymmetric states. This should be taken into account when comparing supergravity and SCFT results using the scheme above, as we are doing here. Alternatively, one should introduce some non-standard counterterms \cite{Genolini:2016sxe,Genolini:2016ecx} that remove the anomaly from the supersymmetric Ward identities, at the price of sacrifying other symmetries. For most of our discussion below this issue will not be important, however we will make explicit comments at the points where it may play a role.

\subsection{Conserved charges}

We next evaluate the one-point functions defined above on the near-boundary solution of Section \ref{sec:near_boundary_sol}, using its Fefferman-Graham form given in Appendix \ref{app:FeffermanGraham}. In order to do so we will not need to make any assumption about regularity of the solution in the interior of the bulk spacetime. Recall that the near-boundary solution depends on the parameters $a_0$, $a_2$, $a_4$, $a_6$, $v$, $\HL$, $H_2$, $H_4$, and that we trade $a_4, a_6, H_4$ for the first integrals $\mathcal{K}_1, \mathcal{K}_2, \mathcal{K}_3$ defined in Section \ref{sec:FirstIntegrals}, which considerably simplifies the expressions. The contractions in Appendix~\ref{app:conditions_qI} are also needed in the computations.

We find that the energy momentum tensor \eqref{Stress_Tensor_QFT_explicit} can be expressed as:
\begin{equation}
\braket{T_{ij}} \diff x^i \, \diff x^j = \braket{T_{tt}} \diff t^2 + \braket{T_{\theta \theta}} \left(\sigma_1^2 + \sigma_2^2 \right) + \braket{T_{\psi \psi}} \sigma^2_3 + 2 \braket{T_{t \psi}} \diff t \, \sigma_3 \ ,
\end{equation}
where the components read:
\begin{align}\label{enmomtensor_explicit}
\braket{T_{tt}} &= \frac{1}{\kappa^2  a_0^2\,  v^4  \ell} \left( \big(\tfrac{1}{9} - \HL^2 - 2 \, \mathcal{K}_1 \big) v^2 - \tfrac{7}{36}  v^4 + \tfrac{89}{864}  v^6 + 2  \HL \big(2 \HL^2 - \HL + 6 \, \mathcal{K}_2\big) \right. \notag \\
& \hspace{2.3cm}   +\tfrac{1}{27} \big(2 - 108 \mathcal{K}_1 + 27 \mathcal{K}_3\big) \Big)\ ,\notag\\[1mm]
\braket{T_{\theta \theta}} &= \frac{\ell}{384\kappa^2  a_0^2} \, \left( 16   ( 16 a_2-5)v^2 + 67 \, v^4 + 288 \, \HL\, (4 \, H_2 + \HL) + 32 - 576 \, \mathcal{K}_1 \right)\ ,\notag
\end{align}
\begin{align}
\braket{T_{\psi \psi}} &= \frac{\ell}{3456\kappa^2  a_0^2} \, \Big(4320 \,v^2 \HL^2  - 480 \left(1 - 18 \, \mathcal{K}_1\right)  v^2 - 24 \, (192 \, a_2-53 ) \, v^4 - 
1117 v^6  \notag \\ 
&\hspace{2.5cm}  + 1728 \, \HL \, (2 \HL^2 - \HL + 6 \, \mathcal{K}_2)  + 32 \, (2 - 108 \, \mathcal{K}_1 + 27 \, \mathcal{K}_3)\Big)\ ,\notag\\[1mm]
\braket{T_{t \psi}} &= \frac{1}{\kappa^2 a_0^2 \, v^2} \left(  \tfrac{1}{27} (v^2-1)^3  -   (v^2-1)\HL^2 -2 \HL^3 - 2 \mathcal{K}_1 (v^2-1) - 6 \HL  \mathcal{K}_2 - \tfrac{1}{2} \mathcal{K}_3  \right)\, .
\end{align}
The trace is:
\be
\braket{T_i{}^{i}} = \frac{3}{\kappa^2 a_0^4}\HL\left( 2 H_2 +\HL \right)\ .
\ee

The non-vanishing components of the electric current \eqref{Holo_current} are:
\begin{align}\label{current_explicit}
\langle j_I^t \rangle &= \frac{-1}{36 \kappa^2 \ell^2 a_0^4 } \left[  \left(54 \,  \mathcal{K}_1 -\big(v^2 - 1\big)^2 +9  \HL^2 \right)  \bar{X}_I + 6 \, \left(9 \mathcal{K}_2 + \big(v^2-1\big)\HL +3 \HL^2  \right) q_I \right] ,\nn\\[1mm]
\langle j_I^\psi \rangle &= \frac{1}{72 \kappa^2\ell^2 a_0^4 v^2} \left[  \left( 4 \big(36 a_2 -5 \big) v^2-36 \, \HL^2 - 216 \, \mathcal{K}_1 + 25 \, v^4 + 4 \right) \bar{X}_I \right. \notag \\ 
&\hspace{3cm}\left. - 12 \, \left(18  \left(H_2 \, v^2 + \mathcal{K}_2 \right) + \HL \big(6 \, \HL+ 5 \, v^2 - 2\big) \right)  q_I \right]\ .
\end{align}

In the limit $\HL=H_2=\mathcal{K}_1=\mathcal{K}_2=\mathcal{K}_3=0\,$, \eqref{enmomtensor_explicit} and \eqref{current_explicit} are consistent with the energy-momentum tensor and current for minimal gauged supergravity solutions presented in \cite{Cassani:2014zwa}.\footnote{We correct an overall sign mistake in the expression for $\langle T_{\psi\psi}\rangle$ appearing in Appendix B of \cite{Cassani:2014zwa}; we thank P. Benetti Genolini for pointing this out. In order to match the current one has to take into account that the relative normalization between the gauge field in \eqref{A_minimal} and the one in \cite{Cassani:2014zwa} is $A^{\rm here} = \frac{2}{\sqrt{3}}A^{\rm there}$. One should also note that a different gauge choice is made, which affects the $\psi$ component of the current.}

The scalar one-point function \eqref{vev_O_general} evaluates to:
\be
\langle\mathcal{O}_I\rangle \,=\, -\frac{3}{\kappa^2a_0^2} \left(2H_2 +\HL\right) q_I\ .
\ee

It is easy to check that the Ward identities \eqref{conservation_current}--\eqref{conformal_Ward} are satisfied with 
\be
 \mathcal{A}^{\rm chiral}_I =  \mathcal{A}^{\rm Weyl} = 0 \ .
 \ee
Moreover the two sides of \eqref{conservation_enmomtensor} actually vanish separately on our background, so the energy-momentum tensor satisfies the standard conservation law $\nabla^{i}\langle T_{ij} \rangle=0$.

Vanishing of both the chiral and Weyl anomalies is a consequence of supersymmetry.  
Indeed both $\mathcal{A}^{\rm Weyl}$ and $\mathcal{A}^{\rm chiral}_I$ must be four-dimensional superconformal invariant Lagrangian built out of background conformal supergravity multiplets. As already discussed, in our holographic setup the latter arise as the asymptotic values of the bulk fields and belong to the Weyl multiplet and $n_V$ vector multiplets. 
It was shown in \cite{Cassani:2013dba,Assel:2014tba} that the respective superconformal invariant Lagrangians vanish on supersymmetric backgrounds of the type studied in this paper, implying that the gauge and conformal Ward identities are satisfied with no anomalous contribution.

Since the holographic electric currents $\langle j^i_I \rangle$ are conserved, we can introduce holographic electric charges $Q_I$ as:
\be
Q_I = \int_{\Sigma_\infty} {\rm vol}_{\Sigma}\, u_i \langle j^i_I \rangle\ ,
\ee
where $u^i$ is a unit timelike vector for the metric on the conformal boundary $\partial M$ and $\Sigma_\infty \subset \partial M$ is a compact, spacelike hypersurface in the boundary.
Using \eqref{Holo_current} it is not hard to show that this is the same as:\footnote{The overall minus sign can be traced back to the fact that our choice of orientation for the bulk and the boundary is such that ${\rm vol}(M)=-\frac{\diff r}{r} \wedge {\rm vol}(\partial M)$.}
\begin{equation}
\label{Holo_charge}
Q_I  = -\frac{1}{\kappa^2} \int_{\Sigma_\infty} { \left( Q_{IJ} \star F^J + \frac{1}{6} C_{IJK} \, A^J \wedge F^K \right) }\ .
\end{equation}
It follows that the holographic electric charges are related to the Page charges \eqref{Page_charges} as:
\be\label{HoloCharge_from_PageCharge}
Q_I = - P_I + \frac{1}{12\kappa^2}\int_{\Sigma_\infty} C_{IJK}A^J \wedge F^K\ .
\ee
The holographic electric charges and the Page charges do not agree due to the different contributions from the Chern-Simons term \cite{Marolf:2000cb}. Since $\mathcal{A}_{\text{chiral}}=0$, both are invariant under small gauge transformations, however they transform under large gauge transformations.  Evaluating either one of the formulae above,
we obtain:
\begin{equation}
\label{Holo_Charge_final}
 Q_I = 
\frac{16\pi^2\ell^2}{\kappa^2} \left[ \left(3\, \mathcal{K}_1  - \tfrac{1}{18} \left(v^2 - 1\right)^2 + \tfrac{1}{2}  \HL^2\right) \bar{X}_I \, + \,  \left(3 \, \mathcal{K}_2 + \tfrac{1}{3} (v^2 - 1) \, \HL + \HL^2 \right ) q_I \right]\,.
\end{equation}
 
Given an asymptotic symmetry of the solution generated by a vector $Z$, we can also define the associated conserved charge
\be\label{holographic_conserved_quantity}
Q_Z = \int_{\Sigma_\infty} \text{vol}_\Sigma \, u_i \left( \,\braket{T^i{}_j}  + A^I_j\, \langle j^i_I\rangle \, \right)Z^j \ ,
\ee
where the term involving $\langle j_I\rangle$ is in general required because the energy-momentum tensor satisfies the modified conservation equation \eqref{conservation_enmomtensor} (although in our background the energy-momentum tensor actually satisfies the standard conservation law and thus in principle we could define conserved quantities just in terms of it).
In particular, the holographic energy and angular momentum may be defined as the charges associated with the vectors $\frac{\partial}{\partial t}$ and $-\frac{\partial}{\partial \psi}$, respectively:
\begin{equation}
\label{Energy}
E = Q_{\frac{\partial}{\partial t}} = \int_{\Sigma_\infty} \text{vol}_\Sigma \, u_i \left( \,\braket{T^i{}_t}  + A^I_t\, \langle j^i_I\rangle \, \right) \ ,
\end{equation}
\begin{equation}
\label{Angular_Momentum}
 Q_{-\frac{\partial}{\partial \psi}} = -\int_{\Sigma_\infty} \text{vol}_\Sigma \, u_i \left( \,\braket{T^i{}_\psi}  + A^I_\psi\, \langle j^i_I\rangle \, \right)\ .
\end{equation}
By using our expressions for the energy-momentum tensor and the electric currents, we find:
\begin{equation}
\label{Energy_computed}
E = \frac{\pi^2\ell^2}{\kappa^2} \left( \frac{16}{9} -\frac{14}{9}v^2 + \frac{19}{36}v^4 -16 \, \HL^2 + \frac{8}{v^2}\mathcal{K}_3 \right) \, ,
\end{equation}
\begin{equation}
\label{Angular_Momentum_computed}
Q_{-\frac{\partial}{\partial \psi}} = \frac{4\pi^2\ell^3}{\kappa^2}\mathcal{K}_3 =J\ ,
\end{equation}
where for the last equality we used \eqref{defK3}. This shows that the holographic angular momentum coincides with the generalized Komar integral~\eqref{Komar_int}.
These results for the electric charges $Q_I$, the energy $E$ and the angular momentum $J$ hold for any AlAdS solution to Fayet-Iliopoulos gauged supergravity satisfying the supersymmetry equations \eqref{eqforH}, \eqref{new_eqfora}. The expressions depend only on the squashing at the boundary $v$, on the scalar source $\HL$ and on the constants $\mathcal{K}_1, \mathcal{K}_2, \mathcal{K}_3$. As explained in Section \ref{sec:FirstIntegrals}, the latter can be fixed by studying how the solution caps off in the interior.

We recall that for our two-parameter family of black hole solutions, the value of $\mathcal{K}_1, \mathcal{K}_2, \mathcal{K}_3$ is given in terms of the near-horizon parameter $\air$ in \eqref{K1_IR}--\eqref{K3_IR}, while we could relate the boundary data $v$ and $\HL$ to the near-horizon parameters $\air$ and $\xi$ only numerically (recall Figures \ref{fig:xi_v_plot}, \ref{fig:H_v_plot}).%\footnote{As a cross-check, we verified that in the minimal limit $\HL = 0$, $\mathcal{K}_2=0$, $\air = \sqrt{\frac{8}{11}}$, the holographic electric charge of our black hole solution is in agreement with the one of \cite{Blazquez-Salcedo:2017ghg} after taking into account the different normalization of the gauge field.}

\subsection{On-shell action and quantum statistical relation}\label{sec:OurOnShAct}

We now proceed to evaluate the renormalized action on our supersymmetric black hole solution.\footnote{For the solutions of \cite{Gutowski:2004ez,Gutowski:2004yv}, the on-shell action was computed in \cite{Liu:2004it}.} This is somewhat formal: a physically more meaningful way to compute the on-shell action of an extremal solution would be to start from a non-extremal generalization having a regular Euclidean section,  evaluate the corresponding on-shell action, and then take the extremal limit. Nevertheless we find it useful to proceed with a direct evaluation of the action on our Lorentzian solution since in addition to exhibiting the cancellation of the large-distance divergences for all asymptotic solutions of Section \ref{sec:near_boundary_sol}, it will lead to a result with a simple physical interpretation. 

We start from the bulk action $\eqref{Bulk_action}$. Using the trace of the Einstein equation \eqref{Einstein_equations} and rewriting the Chern-Simons term by means of the Maxwell equation \eqref{Maxwell_general}, this can be expressed as:
\begin{equation}\label{Sbulk_rewrite}
S_{\text{bulk}} \,=\,  \frac{2}{3 \kappa^2} \int_{M_{r_0}} {  \CV \star 1 } \,-\, \frac{1}{3\kappa^2} \int_{M_{r_0}} { \diff \left(Q_{IJ}\, A^I \wedge \star F^J \right) }\ .
\end{equation}
Since $Q_{IJ} \, A^I \wedge \star F^J$ is globally well-defined and vanishes at the horizon {\it in the chosen gauge}, the second term reduces by the Stokes theorem to an integral over the boundary $\partial M_{r_0}$.  
The same is true for the first term. This can be seen by noticing that using \eqref{eqforf}, the scalar potential \eqref{scalarpot} reads:
\begin{equation}
\label{Scalar_Potential_Explicit}
\CV = -6\, \ell^{-2} \,\bar{X}^I X_I = -6 \,\ell^{-2} f f^{-1}_\text{min}  \ , 
\end{equation}
which implies
\begin{equation}
 \mathcal{V}\star 1 = -12\ell^{-2} \,f^{-1}_\text{min} \, a^3  a^\prime \, \diff t\wedge \diff \rho\wedge\sigma_1\wedge\sigma_2\wedge\sigma_3 = \frac{1}{2} \,\diff\left(a^2 p \,\diff t  \wedge \sigma_1\wedge\sigma_2\wedge\sigma_3\right)\, ,
\end{equation}
where in the last equality we used \eqref{curvature_from_ricci_pot}. 
The integral on $M_{r_0}$ is now trivially performed. Since from the analysis of Section \ref{sec:near_horizon_sol} it follows that $a^2p\to 0$ at the horizon, we obtain that the only contribution is from the upper limit of integration. Thus the bulk supergravity action can be expressed as a term evaluated at $r=r_0$ as:\footnote{The positive orientation on the five-dimensional spacetime is defined by $\diff t\wedge \diff \rho\wedge\sigma_1\wedge\sigma_2\wedge\sigma_3$, while we choose $\diff t\wedge\sigma_1\wedge\sigma_2\wedge\sigma_3$ as the positive orientation on the boundary. As a consequence, the Stokes theorem reads $\int_{M_{r_0}} \diff\omega=-\int_{\partial M_{r_0}}\omega$.}
\begin{equation}
\label{Bulk_action_final}
S_{\text{bulk}} = - \frac{16  \pi^2}{3\kappa^2} \, a^2 p\,|_{r_0} \int{\diff t} + \frac{1}{3\kappa^2} \int_{\partial M_{r_0}} {Q_{IJ}\,A^I \wedge \star F^J }\ .
\end{equation} 
 Using the asymptotic expansion of the $a$ function obtained in Section~\ref{sec:near_boundary_sol}, the first term in \eqref{Bulk_action_final} evaluates to: 
\begin{align}
\label{Bulk_gravity_UV}
- \frac{16 \pi^2}{3\kappa^2} \, (a^2 p)|_{r_0}\! \int\!{\diff t} &\approx - \frac{8\pi^2\ell^2}{\kappa^2}  \bigg[ 4 \,  a_0^4 \, \left(\frac{r_0}{\ell}\right)^4 - \frac{1}{3}  \, (4 c + 3) \, a_0^2 \, \left(\frac{r_0}{\ell}\right)^2 - \frac{32}{9} \, c^2 \, \log \frac{r_0}{\ell} \notag \\ 
&\,\quad+ \frac{1}{36} \, (-128 \, a_2 + 38 \, c + 1) \, c - \HL^2 - 2 \, \mathcal{K}_1 + \frac{3}{32} \bigg] \int\!{\diff t}\,,
\end{align}
where the symbol $\approx$ means that the equality holds up to terms that vanish as $r_0\to\infty$.
The second term in \eqref{Bulk_action_final} is less straightforward. Recalling that $A^I$ is given by \eqref{gauge_field} and  $\star F^I$ by \eqref{Field_strength_star}, we can write:
\be
A^I\wedge \star F^J = \left[2a^3a'f^{-1}X^I \left(f X^J \right)^\prime - \frac{af}{2a^\prime} U^I\!\left(f w^\prime X^J + \left(U^J\right)' \right) \right]\! \diff t\wedge\sigma_{1}\wedge\sigma_{2}\wedge\sigma_{3}\, .
\ee
Both $X^I$ and $U^I$ contain a part proportional to $\bar{X}^I$ and a part proportional to $\bar{Y}^I$, as it is apparent from their expressions \eqref{UI_from_ansatz}, \eqref{scalars_up_from_ansatz}. With the aid of~\eqref{Q_IJ_contractions} we can evaluate the contractions of the different terms with the kinetic matrix $Q_{IJ}$. In this way we arrive at an expression for $Q_{IJ}A^I \wedge \star F^J$ that we expand asymptotically. Doing so we obtain:
\begin{align}
\label{Bulk_gauge_UV}
& \frac{1}{3\kappa^2} \int_{\partial M_{r_0}} { Q_{IJ}\, A^I \wedge \star F^J}  \approx  \notag \\[1mm] 
& - \frac{8\pi^2 \ell^2}{\kappa^2} \bigg[\frac{4}{9} \big(8 \, c^2 + 9 \, \HL^2 \big) \log \frac{r_0}{\ell}  + \frac{2}{9} \!\left(1 + 16 \, a_2 - 12 \, c \right) c + \big(4  H_2 + \HL \big) \HL + 2  \mathcal{K}_1 \bigg] \!\int\!\!{\diff t}\ ,
\end{align}
which concludes our evaluation of the bulk action \eqref{Bulk_action_final}.
In both expressions resulting from \eqref{Bulk_action_final} the parameter $a_4$ has been traded for the Page charge $\mathcal{K}_1$ using  $\eqref{a4_condition}$. 

The Gibbons-Hawking term yields:
\begin{equation}
\label{Gibbons_Hawking_Asymptotic}
S_{\text{GH}} \approx - \frac{8\pi^2 \ell^2}{\kappa^2}\! \left[-16 a_0^4 \left(\frac{r_0}{\ell}\right)^4\! + \Big( 1+  \frac{4}{3}  c \Big) a_0^2 \left(\frac{r_0}{\ell}\right)^2\! + 8  \HL^2  \log \frac{r_0}{\ell} + 8 H_2 \HL + 4 \HL^2 \right] \!\int\!{\diff t} .
\end{equation} 

We finally evaluate the counterterm action~\eqref{Counterterm_action}. This is most easily done using the asymptotic expansion of the supergravity fields given in Appendix~\ref{app:FeffermanGraham}, also recalling some of the contractions in Appendix  \ref{app:conditions_qI} to evaluate the term involving the gauge field. 
We obtain:
\begin{equation}
\label{Counterterms_UV}
S_{\text{ct}} \approx -\frac{8\pi^2\ell^2}{\kappa^2} \left[12 \, a_0^4 \, \left(\frac{r_0}{\ell}\right)^4 - 12 \, \HL^2 \, \log \frac{r_0}{\ell} + \frac{8}{3}c^2 -6 \, \HL \left(2 H_2 + \HL \right) \right] \int{\diff t}\ .
\end{equation}
Notice that as long as $\HL \neq0$, namely as long as the scalar source term is non-vanishing, the counterterm action contains a logarithmic divergence in addition to a power-law divergence. As explained under eq.~\eqref{Weyl_anomaly}, there is no contradiction with the fact that $\mathcal{A}^{\rm Weyl}=0$.

Adding up \eqref{Bulk_gravity_UV}, \eqref{Bulk_gauge_UV}, \eqref{Gibbons_Hawking_Asymptotic}, \eqref{Counterterms_UV} and removing the cutoff, we arrive at our result for the renormalized on-shell action:
\begin{equation}
\label{On-Shell_action_renormalized}
S_{\text{ren}} = - \frac{\pi^2\ell^2}{\kappa^2} \left[\frac{16}{9} - \frac{14}{9} \, v^2 + \frac{19}{36} \, v^4  - 16 \,\HL^2 \right] \int{\diff t}\ .
\end{equation}
This depends only on the squashing at the boundary $v^2$ and on the scalar source term $\HL$. 
The expression is valid for solutions that have the near-boundary behaviour discussed in Section~\ref{sec:near_boundary_sol} and that in addition have no contributions from the lower limit of integration of the bulk action.
We also remark that a priori the final result for the on-shell action depends on large gauge transformations.
The gauge-dependence arises from the Chern-Simons term in the bulk action (or, after using the equations of motion, from the second term in $\eqref{Sbulk_rewrite}$). The appropriate gauge to be used for evaluating the on-shell action may be prescribed by regularity of the  solution. Here we used a gauge condition such that $V^\mu  A^I_\mu = 0$ at the horizon, cf.~eq.~\eqref{our_gauge}, which avoids a divergence in the square norm of the gauge fields. In this gauge, the Killing spinor parameterizing the supersymmetry of the solution is preserved by the vector $V$ given in \eqref{Killing_tpsi} (recall that in Fayet-Iliopoulos gauged supergravity the supersymmetry parameter is charged under $\bar{X}_I A^I$ and therefore the expression for the Killing spinor is gauge-dependent). 
It should be noted that when taking the minimal limit $\HL\to0$, this gauge choice leads to an expression for the on-shell action that is different from the one given in \cite[eq.$\:$(4.13)]{Cassani:2014zwa}. 
Indeed in \cite{Cassani:2014zwa} a different gauge choice was made,\footnote{From \eqref{bdry_AX} we see that the present gauge satisfies $\lim_{r\to\infty}V^\mu  A_\mu^I \bar{X}_I= 1 $, while the gauge chosen in \cite{Cassani:2014zwa} corresponds to $\lim_{r\to\infty}V^\mu  A^I_\mu \bar{X}_I = 1-\frac{2}{3v^2}$.} ensuring that the Killing spinor is instead preserved by the vector $\frac{\partial}{\partial t}$. This was required by global well-definiteness of the spinor in the solitonic geometry studied in that paper, where after a Wick rotation $\frac{\partial}{\partial t}$ generated translations along an $S^1$ of finite size.

We notice that the on-shell action \eqref{On-Shell_action_renormalized} satisfies the simple relation 
\begin{equation}
\label{BPS_relation}
 - \frac{S_\text{ren}}{\Delta_t} = Q_V\ ,
\end{equation}
where we have defined $\Delta_t=\int{\diff t} $ and 
\be
Q_V = E - \frac{2}{\ell v^2}J
\ee
 is the holographic charge associated with the supersymmetric Killing vector \eqref{Killing_tpsi}. 
This relation can be interpreted as a limit of the quantum statistical relation for general AlAdS spacetimes. The latter reads (see~\cite{Papadimitriou:2005ii} for a discussion in the context of holographic renormalization):
\be
\frac{I}{\beta} = E - T \mathcal{S} - \Omega J - \Phi^I Q_I\ ,
\ee
where $I$ is the Euclidean on-shell action, $\mathcal{S}$ is the entropy, $T = 1/\beta$ is the temperature, $\Omega$ is the angular velocity of the horizon measured with respect to a static frame at infinity, and $\Phi^I$ is the electric potential. 
Taking the limit to extremality and considering just the leading order terms, we obtain: 
\be\label{QuantumStatRelT=0}
\frac{I}{\beta} = E - \Omega J - \Phi^I Q_I\ ,
\ee
where now all quantities are evaluated in the extremal solution.

In our setup, the electric potential is:\footnote{Here we are using the definitions of \cite{Papadimitriou:2005ii}, where the electric potential is measured just at the horizon, $\Phi^I = V^\mu A_\mu^I\, |_{\text{hor}}$ and $E$, $J$ are those introduced in \eqref{Energy}, \eqref{Angular_Momentum}. 
In another possible definition, the electric potential also receives a contribution from the gauge field at infinity, $\Phi^I = V^\mu A_\mu^I\, |_{\text{hor}} - V^\mu A_\mu^I\, |_{\infty}$, while $E$ and $J$  are computed just from the energy-momentum tensor (if conserved), without the term involving the gauge field. In any case the combination $E - \Omega J - \Phi^I Q_I$ remains the same.
} 
\be
\Phi^I \equiv V^\mu A_\mu^I\, |_{\text{hor}} = 0\ ,
\ee
while the angular velocity is read from the vector \eqref{Killing_tpsi} and is:
\be
\Omega = \frac{2}{\ell v^2}\ .
\ee
We see that the right hand side of \eqref{QuantumStatRelT=0} is just $Q_V$. It follows that after identifying $- \frac{S_\text{ren}}{\Delta_t}$ with its Euclidean continuation $\frac{I}{\beta}$, we can interpret the relation \eqref{BPS_relation} as the leading order term in the extremal limit of the quantum statistical relation.
Note that the entropy does not appear at this order in the limit to extremality: to see it one should consider the next-to-leading order terms.

The same relation \eqref{BPS_relation} can also be seen as the BPS relation between the holographic charges including the anomalous contribution discussed in \cite{Papadimitriou:2017kzw,An:2017ihs}.

\subsection{Entropy}

The expression for the entropy of our black hole solution follows from the area of the horizon given in \eqref{AreaHorizon}. It is interesting to note that this can be expressed as a simple combination of the Page charges and the angular momentum of the solution. Indeed, using \eqref{K1_IR}--\eqref{K3_IR} into \eqref{AreaHorizon} we arrive at:
\begin{align}\label{RelationEntropyCharges}
\CS = \frac{2\pi}{\kappa^2}\text{Area} &=\frac{8\pi^3\ell^3}{\kappa^2}\sqrt{48\, \mathcal{K}_1^2 - 12\, \mathcal{K}_2^2 - \mathcal{K}_3}\nn\\[2mm]
&= 2\pi\ell\sqrt{\frac{3}{2}C^{IJK}\bar{X}_I P_J P_K - \frac{4\pi^2\ell}{\kappa^2} J }\ .
\end{align}
This is the same relation found in \cite{Kim:2006he} for the asymptotically AdS$_5$ black holes of \cite{Gutowski:2004ez,Gutowski:2004yv}. The fact that the same relation holds here is certainly not surprising, since on the one hand we have seen in Section \ref{sec:near_horizon_sol} that our horizon geometry forms a one-parameter sub-family of the horizon geometry of \cite{Gutowski:2004yv}, and on the other hand all quantities appearing in \eqref{RelationEntropyCharges} can be measured at the horizon (recall the discussion of Section \ref{sec:FirstIntegrals}). 
However, it is important to note that while in the asymptotically AdS$_5$ case the Page charges $P_I$ and holographic charges $Q_I$ essentially coincide because the additional boundary contribution in \eqref{HoloCharge_from_PageCharge} vanishes, in the present asymptotically locally AdS$_5$ case they are different, and we find that the relation  \eqref{RelationEntropyCharges} really involves the Page charges. In other words, this relation does not hold in our solution if the $P_I$ are replaced with the $Q_I$, due to the dependence of the latter  on additional boundary data.
 
Recently, an extremization principle has been proposed \cite{Hosseini:2017mds}, where the expression \eqref{RelationEntropyCharges} for the entropy of the supersymmetric asymptotically AdS$_5$ black holes of \cite{Gutowski:2004ez,Gutowski:2004yv} is reproduced by the Legendre transform of a certain function of chemical potentials that are conjugate to the black hole charges and angular momenta.\footnote{In \cite{Hosseini:2017mds} this principle was also discussed for the supersymmetric AdS$_5$ black holes with two indipendent angular momenta of \cite{Chong:2005hr,Chong:2005da,Kunduri:2006ek}, while in \cite{Hosseini:2018dob} it was extended to AdS$_7$ black holes.} This is particulary appealing as the function of chemical potentials has a close resemblance with the supersymmetric Casimir energy of four-dimensional superconformal field theories (SCFT's) on $S^1 \times S^3$ \cite{Assel:2014paa,Assel:2015nca,Bobev:2015kza} (this relation has been made precise in \cite{Cabo-Bizet:2018ehj}). It is natural to ask whether the same extremization principle would hold for the black hole solution presented in this paper. Our observations above indicate that the same extremization will go through and give the entropy as a result, provided the extremization variables for the function defined in \cite{Hosseini:2017mds} are understood as chemical potentials conjugate to the Page charges $P_I$. The failure of the holographic charges $Q_I$ to reproduce the entropy when they are inserted in \eqref{RelationEntropyCharges} at the place of the $P_I$ may also be related to the choice of supersymmetric scheme discussed at the end of section \ref{sec:HoloRenoFIsugra}.

%%%%%%%%%%%%%%%%%%%%%%%%%%%%%%%%%%%%%%%%%%%%%%%%%%%%%%%%%%%%%%%%%%%%%%%%%%%%%%%%%%%%%%
\section{Conclusions}\label{sec:conclusions}

In this paper we have presented a new two-parameter family of supersymmetric AlAdS$_5$ black hole solutions comprising a squashed $S^3$ at the conformal boundary. 
We have seen that one of the parameters controls the event horizon geometry as well as the angular momentum and the Page electric charges, while the other can be identified with the squashing of the $S^3$ at the boundary. Suppose we fix the former. Then although the squashing at the boundary is arbitrary, the $S^3$ metric flows to a fixed one at the horizon. This is reminiscent of the attractor mechanism for scalar fields in four dimensions. This connection can be made rigorous by reducing along the Hopf fiber of $S^3$, as in the dimensional reduction the component of the metric controlling the size of the Hopf fiber becomes one of the scalar fields involved in the attractor mechanism (see \cite{Hosseini:2017mds} for a related discussion in the case with no squashing).

The fact that the solution depends on one parameter in addition to the squashing deserves some remarks. 
Let us consider for definiteness the $n_V=2$ model that arises as a consistent truncation of type IIB supergravity on $S^5$. In this case a solution carries  energy, one angular momentum (associated with rotation in the $\SU(2)\times\U(1)$ symmetric external space) and three electric charges (associated with $\U(1)^3$ rotations in $S^5$). Supersymmetry imposes one linear relation between these quantities, which would a priori leave us with four independent charges. Already in the solution with no squashing of \cite{Gutowski:2004yv}, however, one obtains just three independent parameters as a second constraint needs to be enforced in order to avoid causal pathologies \cite{Cvetic:2005zi}. 
Given this counting, we could expect that it is possible to obtain a black hole solution controlled by three independent parameters in addition to the squashing at the boundary. One reason why this is not the case in our solution may be that the simplifying ansatz made in Section~\ref{sec:simpl_ansatz} is too restrictive, although it should be noted that it is perfectly compatible with the multi-charge solution of \cite{Gutowski:2004yv}. It would be interesting to see if by relaxing this ansatz more general black holes can be found in the $n_V=2$ model. It is also conceivable, although harder to verify, that the additional solutions break the $\SU(2)\times\U(1)^4$ symmetry in the bulk. In this case five-dimensional Fayet-Iliopoulos gauged supergravity would be a too limited setup and one should rather work in a more general consistent truncation or directly in ten dimensions.

Another interesting avenue for future research will be to extend the study of supersymmetric AlAdS black holes with a deformed boundary done in this paper to other dimensions, the seven-dimensional case being perhaps the most promising.

%%%%%%%%%%%%%%%%%%%%%%%%%%%%%%%%%%%%%%%%%%%%%%%%%%%%%%%%%%%%%%%%%%%%%%%%%%%%%%%%%%%%%%
\section*{Acknowledgments}

We would like to thank Stefano Giusto for useful discussions and especially Dario Martelli for collaboration in the initial stages of this work and comments on the manuscript.

%%%%%%%%%%%%%%%%%%%%%%%%%%%%%%%%%%%%%%%%%%%%%%%%%%%%%%%%%%%%%%%%%%%%%%%%%%%%%%%%%%%%%%
%%%%%%%%%%%%%%%%%%%%%%%%%%%%%%%%%%%%%%%%%%%%%%%%%%%%%%%%%%%%%%%%%%%%%%%%%%%%%%%%%%%%%%
\appendix

%%%%%%%%%%%%%%%%%%%%%%%%%%%%%%%%%%%%%%%%%%%%%%%%%%%%%%%%%%%%%%%%%%%%%%%%%%%%%%%%%%%%%%
\section{Useful contractions}\label{app:conditions_qI}

In this appendix we collect various relations involving the parameters $q_I$. Recall that these must be chosen so that 
\be\label{barXq=0_appendix}
\bar{X}^I q_I= 0\ .
\ee

We start by proving that condition \eqref{cond_on_q} on the $q_I$ implies \eqref{CqqMainText}, \eqref{CXqq_and_Cqqq_MainText}, that we report here for convenience:
\begin{align}\label{CXbarqq}
C^{IJK}\bar{X}_Iq_Jq_K &= - \frac{1}{18}\ ,\\[1mm]
C^{IJK}q_Jq_K &= -\frac{1}{18} \bar{X}^I + \bar{Y}^I\ , \quad\text{where}\quad
\bar{Y}^I = 
C^{IJK} \bar{X}_J q_K \ , \label{Cqq}
\\[1mm]
C^{IJK}q_Iq_Jq_K &= - \frac{1}{18} \label{Cqqq}
  \ .
\end{align}

Using \eqref{cond_on_q}, we can compute
\begin{align}
\bar{Q}^{IJ}q_Iq_J &= 36\bar{Q}^{IJ}\left(\bar{Q}_{IK}- \frac{3}{2}\bar{X}_I\bar{X}_K\right)\left(\bar{Q}_{JL}- \frac{3}{2}\bar{X}_J\bar{X}_L\right)(Cqq)^K(Cqq)^L \nn\\[1mm]
&= 36\left(\bar{Q}_{KL}- \frac{3}{2}\bar{X}_K\bar{X}_L\right)(Cqq)^K(Cqq)^L\,.
\end{align}
With the aid of \eqref{Qmatrix}, \eqref{QmatrixInverse}, \eqref{barXq=0_appendix}, this can be rewritten as
\be
(C\bar{X}qq) = 3 \,  C_{KLI}\bar{X}^I(Cqq)^K(Cqq)^L - 18\, (C\bar{X}qq)^2\ .
\ee 
The property \eqref{3Cinto1C} of the $C_{IJK}$ tensor and again \eqref{barXq=0_appendix} imply that the first term in the right hand side vanishes, leaving us with
\be
(C\bar{X}qq) = -18(C\bar{X}qq)^2\,,
\ee
which is the first in \eqref{CXbarqq}. Here we are assuming $(C\bar{X}qq)\neq0$; indeed $(C\bar{X}qq)=0$ would imply $q_I=0$. This follows from the fact that $(C\bar{X}qq)=0$ can also be written as $\bar{Q}^{IJ}q_Iq_J=0$, which since $\bar{Q}$ is non-degenerate implies $q_I=0$.

We can now return to condition \eqref{cond_on_q}, which using \eqref{CXbarqq} becomes
\begin{equation}
\label{W_parallel_new}
\bar{Q}_{IJ} \left(Cqq\right)^J = - \frac{1}{12} \bar{X}_I - \frac{1}{6} q_I\ .
\end{equation}
Multiplying by $\bar{Q}^{-1}$ and using \eqref{QmatrixInverse} we obtain \eqref{Cqq}.
Upon contraction with $q_I$ the latter implies \eqref{Cqqq}. This concludes our proof of \eqref{CXbarqq}--\eqref{Cqqq}.

\medskip

We next report some contractions between the tensor $C_{IJK}$ and the constant vectors $\bar{X}^I$, $\bar{Y}^I$, that we repeatedly use in the computations in the main text. These can be verified with manipulations similar to those described above. 

The $\bar{Y}^I$ vector is orthogonal to $\bar{X}_I$ and its contraction with $q_I$ is fixed such that:
\begin{align}
& \bar{X}_I \bar{Y}^I = 0 \notag \\
& q_I \, \bar{Y}^I = - \frac{1}{18}\ .
\end{align}
Recalling \eqref{QmatrixInverse}, it can also be useful to record that:
\begin{equation}
\bar{Y}^I = - \frac{1}{6} \, \bar{Q}^{IJ} \, q_J\ .
\end{equation}
Furthermore we have the following contractions:
\begin{align}\label{contractions_CXY}
& C_{IJK} \, \bar{X}^J \bar{X}^K = 6\bar X_I\ ,\notag \\
& C_{IJK} \, \bar{X}^J \, \bar{Y}^K = \frac{1}{3} q_I\ , \notag \\
& C_{IJK} \, \bar{Y}^J \, \bar{Y}^K = - \frac{1}{54} \bar{X}_I - \frac{1}{27} q_I\ , \notag \\
& C_{IJK} \, \bar{X}^I \, \bar{Y}^J \, \bar{Y}^K = - \frac{1}{54}\ , \notag \\
& C_{IJK} \, \bar{Y}^I \, \bar{Y}^J \, \bar{Y}^K = \frac{1}{486} \ .
\end{align}

The following additional contractions involving the matrix $Q_{IJ}$ (rather than its determination $\bar Q_{IJ}$ on the AdS$_5$ vacuum appearing in previous formulae) will be useful when evaluating some terms of the on-shell action in Section~\ref{sec:OurOnShAct}:
\begin{align}
\label{Q_IJ_contractions}
& Q_{IJ} \, \bar{X}^I = \left(\frac{9}{2} \left(f \, f^{-1}_\text{min} \right)^2 -3 \, \CA \right) \bar{X}_I + \left( \frac{9}{2} f^2 \, f^{-1}_\text{min} \, \frac{H^\prime}{a^3 \, a^\prime}  - \frac{3}{2} \, \CB \,  \right) \, q_I\ , \notag \\
& Q_{IJ} \, \bar{Y}^J = \left(- \frac{1}{4} \, f^2 f^{-1}_{\text{min}} \, \frac{H^\prime}{a^3 \, a^\prime}  + \frac{1}{12} \, \CB \right) \bar{X}_I + \left( - \frac{1}{4} f^2 \, \left( \frac{H^\prime}{a^3 \, a^\prime} \right)^2 - \frac{1}{6}  \CA + \frac{1}{6}  \CB \right) q_I \ ,
\end{align}
where $\CA$ and $\CB$ are defined as
\begin{equation}
X^I = \CA \, \bar{X}^I + 9 \, \CB \, \bar{Y}^I
\end{equation}
and thus recalling \eqref{scalars_up_from_ansatz} read:
\begin{align}
& \CA = f^2\left[f_{\text{min}}^{-2}  -\frac{1}{4} \left(\frac{H^\prime} {a^3 \, a^\prime} \right)^{\!2\,}\right] \ ,\notag \\[1mm]
& \CB = f^2  \left[f_{\text{min}}^{-1} + \frac{H^\prime} {2a^3 \, a^\prime} \right]\frac{H^\prime} {a^3 \, a^\prime}\ .
\end{align}

%%%%%%%%%%%%%%%%%%%%%%%%%%%%%%%%%%%%%%%%%%%%%%%%%%%%%%%%%%%%%%%%%%%%%%%%%%%%%%%%%%%%%%
\section{Near-boundary solution in Fefferman-Graham form}\label{app:FeffermanGraham}

In this appendix we give some more details on the construction of the general near-boundary solution of Section \ref{sec:near_boundary_sol} and we cast it in Fefferman-Graham form.  This will confirm that the solution is Asymptotically locally AdS and provide information on the role of the different parameters in determining the source and expectation values for the field theory operators dual to our supergravity fields.

We will keep setting the AdS radius $\ell =1$ and use the coordinates $(t,\, \psi)$ introduced in the main text. We recall that these are related to the previous coordinates $(y,\,\hat\psi)$ as:
\be
\label{change_psi_appendix}
y = t \ ,\notag
\qquad\qquad \hat{\psi} = \psi + \chi \, t \ , \quad \text{where}\quad \chi = \frac{2}{4c-1}\ ,
\ee
and that the form of the five-dimensional metric and gauge fields in these coordinates is:
\begin{equation}
\diff s^2 = g_{\rho \rho} \diff  \rho^2 + g_{\theta \theta} (\sigma_1^2 + \sigma_2^2) + g_{\psi \psi} \sigma_3^2 + g_{tt} \diff t^2 + 2 g_{t \psi} \, \sigma_3 \, \diff t\ .
\end{equation}
\begin{equation}
A^I = A^I_t \, \diff t + A^I_\psi \, \sigma_3\ .
\end{equation}
Starting from \eqref{metric},  \eqref{gauge_field} and implementing the change of coordinates, one finds that the respective components take the form:
\begin{align}
 g_{\rho \rho} &= f^{-1}\ , \qquad \qquad g_{\theta \theta} = f^{-1} a^2\ , \qquad \qquad g_{\psi \psi} = - f^2  w^2 + f^{-1} (2 a a^\prime)^2\ , \notag \\[1mm]
 g_{tt} &= -f^2 (1 + \chi \, w)^2 + \chi^2 f^{-1} (2 a a^\prime)^2 \ ,\ \ \qquad g_{t \psi} = -f^2 (1 + \chi \, w) \, w + \chi \, f^{-1} (2 a a^\prime)^2\ ,
\end{align}
\be
A^I_t  = \left(\, f + \chi \, f \, w \right) X^I  + \chi \, U^I \ ,\qquad\qquad
A^I_\psi  = f \, w \, X^I + U^I \ .
\ee

In this appendix we present the asymptotic solution for $\rho\to\infty$.
The large-$\rho$ expressions for $a$ and $H$ have been given in eqs.~\eqref{UVsola}, \eqref{UVsolH}. Using \eqref{f_from_a_ansatz} we obtain for~$f$:
\begin{align}
f &=  1 + \left( \frac{1 + 16 \, a_2 + 4 c}{12} + \frac{4 c}{3} \rho \right) \frac{e^{-2 \rho}}{a_0^2} \notag\\[1mm]
&+ \bigg[ \frac{1} {144} \Big(1 -128 a_2^2+  96 a_2\, c + 8a_2 + 24c - 80 c^2  + 18 \left(8 H_2^2 + 12 H_2 \HL +9 \HL^2 \right)\Big)\notag\\[1mm]
& + \frac{1}{18}\Big( (1-32 a_2 + 12 c) c+ 9 \HL (4 H_2 + 3 \HL)\Big) \rho  + \frac{9  \HL^2 - 8 \, c^2}{9} \, \rho ^2 \bigg] \frac{e^{-4 \rho}}{a_0^4} + \CO(e^{-5 \rho}) \ .
\end{align}
Note that $f \to 1$ as $\rho\to \infty$. Eq.~\eqref{w_from_a_ansatz} gives for $w$: 
\begin{align}
w = & -2 a_0^2\, e^{2 \rho} + \frac{1}{2} + 4 a_2 -2 c + 4 c \rho + \frac{1}{48} \bigg[-352 \, a_2^2 + 32 \, a_2 \, (5 c-1) + 192 \, a_4  \notag \\
& + 8 \, c \, (2-3 c) -1  +18 \left(8 \, H_2^2 + 8 H_2 \, \HL + 3 \, \HL^2 \right) + \notag \\
& \Big(80  \, c \, ( c - 12 \, a_2)+72 \, \HL (6 \, H_2 +5 \HL ) \Big) \, \rho \,  + \left(216 \, \HL^2 - 480\, c^2\right) \rho ^2 \bigg] \frac{e^{-2 \rho}}{a_0^2} +\CO(e^{-3\rho})\ .
\end{align}
Using these expressions we can construct the asymptotic expansion of the supergravity fields. 
The leading order terms have already been given in the main text. In the following we present the needed subleading terms after turning the asymptotic solution in Fefferman-Graham form. This is equivalent to show that the solution is AlAdS$_5$.

The general  Fefferman-Graham form of the metric is: 
\begin{equation}
\label{FGmetric}
\diff s^2 = \frac{\diff r^2}{r^2} + h_{ij} (x,r) \, \diff x^i \, \diff x^j\ ,
\end{equation}
where $r$ is a radial coordinate, $x^i$ are coordinates on the hypersurfaces at fixed $r$. The induced metric $h_{ij}$ on such hypersurfaces can be expanded for $r \to \infty$ as:
\begin{equation}
\label{FGmetricomponents}
h_{ij}(x,r) \,=\, r^2 \bigg[ h_{ij}^{(0)} + \frac{h_{ij}^{(2)}}{r^2} + \frac{h_{ij}^{(4)} + \tilde{h}_{ij}^{(4)} \, \log{r^2} + \tilde{\tilde{h}}_{ij}^{(4)} \, \left(\log{r^2}\right)^2}{r^4} + \dots \bigg]\ ,
\end{equation}
where all terms in the expansion depend on the transverse coordinates $x^i$ only.
The Maxwell field, for which the radial gauge $A^I_r = 0$ is assumed, reads:
\begin{equation}
\label{FGgauge}
A^I(x,r) \,=\, A^{I \, (0)} + \frac{A^{I \, (2)} + \tilde{A}^{I \, (2)} \, \log{r^2}}{r^2} + \dots\ .
\end{equation}
Our scalar fields have mass $m^2\ell^2 = -4$ and are thus dual to SCFT scalar operators of conformal dimension $\Delta=2$. The corresponding Fefferman-Graham expansion is (see e.g.~\cite{Bianchi:2001kw}):
\begin{equation}
\label{FGscalars_up}
X^I = \, \bar{X}^I + \frac{\phi^{I \, (0)} + \tilde{\phi}^{I \, (0)} \, \log{r^2}}{r^2} + \frac{\phi^{I \, (2)} + \tilde{\phi}^{I \, (2)} \, \log{r^2} + \tilde{\tilde{\phi}}^{I \, (2)} \, \left(\log{r^2}\right)^2}{r^4} + \dots \ .
\end{equation}

In order to set our five-dimensional metric \eqref{metric} in the form \eqref{FGmetric}, we need to transform the coordinate $\rho$ into the   Fefferman-Graham coordinate $r$ by imposing:
\begin{equation}
f^{-1/2}(\rho) \, \diff \rho = \frac{\diff r}{r}\ .
\end{equation} 
Solving this equation at large $\rho$, we find that the asymptotic change of coordinate is:
\begin{align}
a_0^2 r^2 = \, & a_0^2 \, e^{2 \rho } + \frac{16 a_2 + 12 c +1}{24} + \frac{2c}{3} \, \rho \notag \\ 
&\!\! +\! \bigg[ \frac{1}{2304}\bigg(\! -768 a_2^2+128 a_2 c+8 c (13-30 c)+3 +  72 \left(8 H_2^2 + 16 H_2 \HL + 13 \HL^2\right)\! \bigg)  \notag  \\
&\!\! + \frac{c \, (c-12 a_2)+9 \, \HL \, (H_2+ \HL)}{18} \, \rho + \bigg(\frac{\HL^2}{4}-\frac{c^2}{3}\bigg) \, \rho ^2 \bigg]\frac{e^{-2 \rho}}{a_0^2} + \CO(e^{-3\rho})\ .
\end{align}

Employing the coordinates $(t,r,\theta,\phi,\psi)$, the five-dimensional metric of our solution reads: 
\begin{equation}
\diff s^2 = \frac{\diff r^2}{r^2} + h_{\theta \theta} (\sigma_1^2 + \sigma_2^2) + h_{\psi\psi}\, \sigma_3^2 + h_{tt}\, \diff t^2 + 2 \, h_{t \psi} \, \diff t \, \sigma_3\ ,
\end{equation} 
where the components $h_{\theta \theta}$, $h_{\psi\psi}$, $h_{tt}$ and $h_{t \psi}$ only depend on $r$ and have an expansion of the form \eqref{FGmetricomponents}, with the coefficients being:
\begin{align}
& h^{(0)}_{\theta \theta} = a_0^2 \, , \quad h^{(2)}_{\theta \theta} = - \frac{3 + 20c}{24} \, , \quad \tilde{h}^{(4)}_{\theta \theta}= \frac{4 \, c (1 - 4 c)+3 \, \HL ( \HL - 4 H_2)}{24 a_0^2} \,, \quad \tilde{\tilde{h}}^{(4)}_{\theta \theta} = -\frac{\HL^2}{8 a_0^2}\ , \notag \\[2mm] 
& h^{(4)}_{\theta \theta} = \frac{1024 a_2^2-384 a_2 c+1536 a_4 + 8 c (74 c-15) - 1 - 24 \left(40 H_2^2 + 64 H_2 \HL +49 \HL ^2\right)}{768 a_0^2} \ ,
\end{align}
\begin{align}
& h^{(0)}_{\psi \psi} = a_0^2 (1 - 4c) \ , \qquad h^{(2)}_{\psi \psi} = \frac{(1-4c)(28 c -3)}{24} \ , \qquad \tilde{\tilde{h}}^{(4)}_{\psi \psi} = \frac{(4 c-1) \HL^2}{8 a_0^2}\ , \notag \\[1mm]
& h^{(4)}_{\psi \psi} = \frac{1}{62208 \, a_0^2}\bigg[3 (-75 + 4608 a_4 - 995328 a_6 + 9604 c) + 
16 \Big(-144 a_2 \big(3 + 4 a_2 (3 + 64 a_2)  \notag \\ 
&\quad \qquad + 2064 a_4\big) + 24 \big((391 - 208 a_2) a_2 + 336 a_4\big) c - 9 (1157 + 2976 a_2) c^2 + 
13420 c^3 \Big) \notag \\
&\quad \qquad+ 82944 H_4 (12 H_2 + 19 \HL) - 
1728 (12 H_2 + 11 \HL) (24 H_2^2 + 48 H_2 \HL + 11 \HL^2) \notag \\
&\quad \qquad + 216 \Big(-8 (9 + 576 a_2 + 268 c) H_2^2 - 
16 (-17 + 304 a_2 + 308 c) H_2 \HL \notag \\
&\quad \qquad + (871 + 2944 a_2 - 
3308 c) \HL^2\Big)\bigg]\ , \notag \\[2mm]
& \tilde{h}^{(4)}_{\psi \psi}  = -\frac{1}{24 a_0^2}(4 c-1) \left(8 \, c \, (4 c-1) - 3 \HL (4 H_2 + 5 \HL)\right) \ ,
\end{align}
\begin{align}
& h^{(0)}_{t \psi} \,=\, h^{(2)}_{t \psi} \,=\,\tilde{h}^{(4)}_{t \psi} \,=\, \tilde{\tilde{h}}^{(4)}_{t \psi} \,=\, 0\ , \notag \\[2mm]
& h_{t \psi}^{(4)} = -2 h_{\theta \theta}^{(4)} - 2 \frac{h_{\theta \theta}^{(0)}}{h_{\psi \psi}^{(0)}} h_{\psi\psi}^{(4)} + \frac{128 a_2 (4 c-1) + 8 c (38 c + 1 ) - 5 - 96 (2 H_2 + \HL)^2}{192 a_0^2} \ ,
\end{align}
\begin{align}
& h^{(0)}_{tt} = - \frac{4 \, a_0^2}{1 - 4c}, \quad h^{(2)}_{tt} = - \frac{4c +3}{6(1-4c)} \, , \quad \tilde{h}^{(4)}_{tt} = -\frac{\HL (4 H_2 + 5 \HL)}{2 a_0^2 (4 c-1)} \, , \quad \tilde{\tilde{h}}^{(4)}_{tt} = \frac{\HL^2}{2 a_0^2 (1-4 c)}\ , \notag \\[2mm]
& h_{tt}^{(4)} = 8\, \frac{h_{\theta \theta}^{(0)}}{h_{\psi \psi}^{(0)}}\, h_{\psi \psi}^{(4)} + 4 \left( \frac{h_{\theta \theta}^{(0)}}{h_{\psi \psi}^{(0)}}\right)^2 h^{(4)}_{\psi \psi}- \frac{1}{48 h^{(0)}_{\psi \psi}} \left[3 + 2 \left(1 - \frac{h_{\psi \psi}^{(0)}}{h_{\theta \theta}^{(0)}} \right) + 11  \left(1 - \frac{h_{\psi \psi}^{(0)}}{h_{\theta \theta}^{(0)}} \right)^2 \, \right]  \notag \\[2mm]
& \quad  \quad \quad -\frac{2 (2 H_2+ \HL )^2}{a_0^2 (4 c-1)}\ .
\end{align} 
The  terms at the leading and next-to-leading orders are identical to those found in \cite{Cassani:2014zwa} for minimal gauged supergravity (see Appendix A of that paper for a comparison), while at the following order the backreaction of the fields in the supergravity vector multiplets, controlled by $\HL$, $H_2$ and $H_4$, deforms the metric. According to a standard holographic analysis, the free terms of the metric are found in $h^{(0)}$ and $h^{(4)}$, which correspond to the source and the expectation value for the energy-momentum tensor of the dual SCFT, respectively. Given the present setup, five free parameters are expected in the metric~\cite{Cassani:2014zwa}, and one can see that the free parameters $a_0$, $c$, $a_2$, $a_4$, $a_6$ indeed appear in the expressions above for $h^{(0)}$ and $h^{(4)}$. 

We then turn to the scalar fields. We find that
they take the form \eqref{FGscalars_up}, with the expansion coefficients:
\begin{align}
\tilde{\phi}^{I \, (0)} & = \frac{9 \, \HL \, \bar{Y}^I}{a_0^2}\ , \notag \\
\phi^{I \, (0)} & = \frac{9 \, \big(2 H_2 + \HL \big) \bar{Y}^I}{a_0^2}\ , \notag \\
\phi^{I \, (2)} & = \frac{ \big(2 H_2 + \HL \big)^2 \bar{X}^I + \big(3 H_2 (4 c + 48 \HL +3) + 9  \HL (-4 c + 10 \HL +1) + 72 H_2^2 \big) \bar{Y}^I }{4 \, a_0^4} \ ,\notag \\
\tilde{\phi}^{I \, (2)} & = \frac{4 \HL \big(2 H_2 + \HL \big) \bar{X}^I + 3 \HL \big(4 c + 48 (H_2 + \HL) + 3\big) \bar{Y}^I}{8 a_0^4} \ ,\notag \\
\tilde{\tilde{\phi}}^{I \, (2)}& = \frac{\HL^2  \bar{X}^I + 18 \, \HL^2 \, \bar{Y}^I}{4 a_0^4} \ .
\end{align}
%For completeness we also provide the expansion of the scalars with lower index, $X_I$, which is easily obtained from \eqref{XIdown_from_XIup}. This takes the form
%\begin{equation}
%\label{FGscalars}
%X_I = \, \bar{X}_I + \frac{\phi_I^{(0)} + \tilde{\phi}_I^{(0)} \, \log{r^2}}{r^2} + \frac{\phi_I^{(2)} + \tilde{\phi}_I^{(2)} \, \log{r^2} + \tilde{\tilde{\phi}}_I^{(2)} \, \left(\log{r^2}\right)^2}{r^4} + \dots 
%\end{equation}
%and the coefficients read
%\begin{align}
%\phi_I^{(0)} & = \frac{\left(2 H_2 + \HL \right) q_I }{a_0^2}\ , \notag \\
%\tilde{\phi}_I^{(0)} & = \frac{\HL q_I}{a_0^2}\ , \notag \\
%\phi_I^{(2)} & = \frac{3 (2 H_2 + \HL)^2 \bar{X}_I + \left(24 \HL (H_2 + \HL) + (3 + 4 c) H_2 + 3 (1 - 4 c) \HL \right) q_I}{12 a_0^4} \ ,\notag \\
%\tilde{\phi}_I^{(2)} & = \frac{12 \HL (2 H_2 + \HL) \bar{X}_I + \HL (3 + 4 c + 24 \HL) q_I}{24 a_0^4}\ , \notag \\
%\tilde{\tilde{\phi}}_I^{(2)}& = \frac{\HL^2 \bar{X}_I}{4 a_0^4}\ .
%\end{align}
The free coefficients are  $\tilde{\phi}^{I\,
(0)}$ and $\phi^{I\,(0)}$, corresponding to the source and the expectation value of the dual scalar operator, respectively, and being controlled by the free parameters $\HL$ and $H_2$.
The expansion of the scalars with a lower index, $X_I$, is easily obtained from the one of $X^I$ using \eqref{XIdown_from_XIup}.

We finally examine the gauge fields. One can write them in the general form \eqref{FGgauge}, with the coefficients being:
\begin{align}
& A^{I \, (0)}_t = \frac{(4 c-3) \bar{X}^I - 108\,  \HL \, \bar{Y}^I }{3 (4 c-1)}\ , \qquad\qquad \tilde{A}^{I \, (2)}_t = 0 \ ,\notag \\
& A^{I \, (2)}_t = \frac {1} {72 a_0^2 (4 c-1)} \bigg[ \Big(-5 + 256 a_2^2 + 384 a_4 + 32 a_2 (-2 + 5 c) + 
8 c (-4 + 29 c)  \notag \\
& - 18 \big(8 H_2^2 + 24 H_2 \HL + 21 \HL^2 \big)\Big) \bar{X}^I + 
432 \Big[12 H_4 - 2 \big(H_2 + 4 a_2 H_2 + \HL - 2 (a_2 + c) \HL\big)  \notag \\
& - 3 \big(4 H_2^2 + 8 H_2 \HL + 5 \HL^2\big) \Big] \bar{Y}^I \bigg]\ ,  
\end{align}
\begin{align}
A^{I \, (0)}_\psi & = -\frac{4}{3} c \bar{X}^I - 18 \, \HL \, \bar{Y}^I , \qquad \tilde{A}^{I \, (2)}_\psi = \frac{(1 - 4 c) (2 \, c \, \bar{X}^I + 27 \, \HL \, \bar{Y}^I) }{6 a_0^2 }\ ,\\
A^{I\, {(2)}}_\psi & = \frac{1}{144 a^2_0} \bigg[ \Big(1 + 256 a_2^2 + 384 a_4 + 32 a_2 (1 - 7 c) + 8 c (-4 + 17 c)  \notag \\
& - 
18 (8 H_2^2 + 24 H_2 \HL + 21 \HL^2) \Big) \bar{X}^I + 
216 \Big(24 H_4 - 
2 (-1 + 8 a_2 + 12 c) H_2 - \HL  \notag \\
& + 8 a_2 \HL - 4 c \HL - 
6 (4 H_2^2 + 8 H_2 \HL + 5 \HL^2) \Big) \bar{Y}^I \bigg]\ .
\end{align}
The equations of motion leave both $A^{I\,(0)}$ and $A^{I \, (2)}$ undetermined in the  Fefferman-Graham expansion of the gauge field, however supersymmetry relates both of them to the metric and the scalar fields. Indeed we find that the only free parameter appearing in the gauge field and not already entering in $h^{(0)}$, $h^{(2)}$, $\phi^{I\,(0)}$, $\tilde{\phi}^{I\,
(0)}$ is $H_4$, which appears in the part of $A^{I \, (2)}$ that is aligned along $\bar{Y}^I$. 

The further subleading coefficients in the Fefferman-Graham expansion of the supergravity fields do not contain any new free parameter and are fully determined by the terms displayed above. 
 As a cross-check of the whole construction, we have verified that the supergravity equations of motion are satisfied up to the first few non-trivial orders.

To summarize, we find that our near-boundary supersymmetric solution can be cast in Fefferman-Graham form and is thus AlAdS$_5$. The source terms $h^{(0)}$, $A^{I\,(0)}$ and $\tilde \phi^{(0)\,I}$ depend on the free parameters $a_0,c$ and $\HL$.
 In particular, $a_0$ and $c$ determine the boundary metric $h^{(0)}$ and the part of $A^{I\,(0)}$ along $\bar{X}^I$,  
 while $\HL$ fixes $\tilde \phi^{I\,(0)}$. The parameters $a_0$, $c$ and $\HL$ together also determine the part of the boundary gauge field $A^{I\,(0)}$ along $\bar{Y}^I$.  
The terms $h^{(4)}$, $A^{I\,(2)}$, $\phi^{I\,(0)}$, related to dual field theory one-point functions, also depend on $a_2$, $a_4$, $a_6$, $H_2$, $H_4$. Their expressions above simplify slightly if $a_4$, $a_6$, $H_4$ are traded for the first integrals $\mathcal{K}_1$, $\mathcal{K}_2$, $\mathcal{K}_3$ introduced in Section \ref{sec:FirstIntegrals} by using \eqref{a4_condition}--\eqref{a6_condition}.

%%%%%%%%%%%%%%%%%%%%%%%%%%%%%%%%%%%%%%%%%%%%%%%%%%%%%%%%%%%%%%%%%%%%%%%%%%%%%%%%%%%%%%
\bibliographystyle{JHEP}
\bibliography{5dBHsquashed}

\end{document}